\documentclass[aps,pre,superscriptaddress,twocolumn,10pt]{revtex4-2}

\usepackage{amsmath}
\usepackage{amssymb}
\usepackage{amsthm}
\usepackage{bbm}
\usepackage{color}
\usepackage{graphicx}[floatfix]
\usepackage{hyperref}
\usepackage[utf8]{inputenc}
\usepackage[T1]{fontenc}
\usepackage{physics}
\usepackage{times}
\usepackage{ulem}

\hypersetup{
    colorlinks=true,linkcolor=blue,citecolor=blue,
    filecolor=blue,urlcolor=blue,breaklinks=true
}

\RequirePackage{xcolor}

\newcommand{\rev}[1]{{\color{red}{#1}}}

\newcommand{\orcid}[1]{\href{https://orcid.org/#1}{\includegraphics[width=7pt]{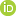}}}

\theoremstyle{definition}

\begin{document}

\title{Unitary description of the Jaynes-Cummings model under fractional-time dynamics}

        \author{Danilo Cius\orcid{0000-0002-4177-1237}}
        \email{danilocius@gmail.com}
        \affiliation{
        	Departamento de F\'{i}sica Matem\'{a}tica, 
        	Instituto de F\'{i}sica da Universidade de S\~{a}o Paulo, 
        	05508-090 S\~{a}o Paulo, Brazil
        }


\begin{abstract}
The time-evolution operator corresponding to the fractional-time Schrödinger equation is nonunitary because it fails to preserve the norm of the vector state in the course of its evolution. However, in the context of the time-dependent non-Hermitian quantum formalism applied to the time-fractional dynamics, it has been demonstrated that a unitary evolution can be achieved for a traceless two-level Hamiltonian. This is accomplished by considering a dynamical Hilbert space embedding a time-dependent metric operator concerning which the system unitarily evolves in time. This allows for a suitable description of a quantum system consistent with the standard quantum mechanical principles. In this work, we investigate the Jaynes-Cummings model in the fractional-time scenario taking into account the fractional-order parameter $\alpha$ and its effect in unitary quantum dynamics. We analyze the well-known dynamical properties, such as the atomic population inversion and the atom-field entanglement, when the atom starts in its excited state and the field in a coherent state.
\end{abstract}

\maketitle

\section{Introduction}
\label{sec:intro}

Many developments at the interface of physics and fractional calculus have been drawing attention. The cornerstone of this approach lies in substituting the $n$-order derivative $\partial_{z}^{n}f(z)$ by fractional-order differential operators denoted as $\mathcal{D}^{\alpha}_{z}f(z)$, which represents fractional derivatives of order $\alpha$ acting on the function $f(z)$ with respect to the variable $z$. The specific definition of the operator depends on the underlying mathematical functions involved \cite{podlubny1998fractional}. Interesting fractional calculus applications appear in statistical physics, particularly in the context of continuous-time random walks (CTRW) to model transport phenomena. CTRW offers a versatile framework for modeling anomalous diffusion processes, often observed in complex systems with memory effects and non-local interactions. 
In the spatial domain, the fractional derivative leads to the emergence of Lévy flights \cite{chaves1998}. These are characterized by jumps with power-law distributed step lengths, resulting in a diffusion process with long-tailed probability densities. 
Conversely, incorporating a fractional derivative in the temporal domain leads to subdiffusive behavior \cite{metzler2000random}. This signifies slower-than-expected diffusion, often observed in systems with heterogeneous environments or trapping mechanisms. 
The choice of which domain (spatial or temporal) to introduce the fractional derivative depends on the specific physical mechanisms governing the transport process. 

A branch of quantum physics rooted in fractional calculus has emerged as a powerful framework for understanding the behavior of quantum systems with non-local, non-Markovian, and long-range interactions. This burgeoning field encompasses diverse areas such as Lévy flights over quantum paths \cite{laskin2007}, optics \cite{longhi2015fractional,huang2017beam}, $\mathcal{PT}$-symmetric systems \cite{zhang2016pt}, the nonlinear variable-order time fractional Schrödinger equation \cite{heydari2019cardinal}, disorder in the vibrational spectra \cite{stephanovich2022}, time-dependent (TD) quantum potentials \cite{gabrick2023}, and anomalous diffusion in a three-level system \cite{lenzi2023}. An experimental demonstration by Wu \textit{et al.} \cite{wu2010} investigated spontaneous emission from a two-level atom in anisotropic one-band photonic crystals. They elegantly employed fractional calculus to resolve an unphysical bound state anomaly arising when the resonant atomic frequency deviates from the photonic band gap. This anomaly, characterized by an infinitely long lifetime, vanishes when the emission peak aligns with the band gap \cite{fujita2005}. 

In this realm, states are described by the fractional Schrödinger equation (FSE), originally proposed by Laskin. Unlike the conventional Schrödinger equation, Laskin's formulation replaces the standard second-order spatial derivative with a fractional Laplacian operator, which is based on the Riesz derivative \cite{laskin:00a,laskin:00b,wei:16,laskin:16}. This modification enables the FSE to model non-local interactions and memory effects, essential for understanding quantum systems' transport phenomena. Furthermore, Naber \cite{naber:04} has proposed a fractional-time Schrödinger equation (FTSE) assuming the Caputo fractional derivative in the place of the ordinary time derivative in such a way that the equation is written as
\begin{equation}
\label{FTSE}
    i^{\alpha} \hbar_{\alpha} {}^{~\text{C}} _{~0}\mathcal{D}_t^{\alpha}\vert\Psi_{\alpha}(t)\rangle
    =
    \hat{H}_{\alpha}\vert\Psi_{\alpha}(t)\rangle \text{,}
\end{equation}
with
\begin{equation}
\label{FD}
{}^{~\text{C}}_{~0}\mathcal{D}_t^{\alpha}(\cdot) =  \int_{0}^{t}d\tau\frac{~(t-\tau)^{-\alpha}}{\Gamma(1-\alpha)~}\frac{d }{d\tau}(\cdot) ,
\end{equation}
defining the fractional Caputo derivative for  $\alpha\in [0,1)$. $\hat{H}_{\alpha}$ represents the fractional Hamiltonian, and $\hbar_{\alpha}$ the fractional-Planck constant used as a scale factor (see  \cite{naber:04}), and therefore, we may consider all the variables and Planck constant in Eq. (\ref{FTSE}) as dimensionless quantities. In Ref. \cite{naber:04}, it is argued that the imaginary unit is raised to the same power as the time coordinate by performing a Wick
rotation. More details about this issue are discussed in \cite{iomin2009fractional}. 
FTSE solutions have been investigated in many settings, including the fractional dynamics of free particles \cite{naber:04} and particles under the influence of $\delta$ potentials \cite{lenzi2013time}.
A mathematical correspondence between the FTSE and the fractional-time diffusion equation \cite{naber:04,iomin2009fractional,lenzi2013time}, viewed as describing a non-Markovian process. 
Also, a connection between classical geometric diffusion and quantum dynamics is elucidated in Ref. \cite{iomin2020}, wherein continuous-time quantum walks are represented as quantum analogs of turbulent diffusion within comb geometry. 

Recent advances in quantum information science, driven by theoretical and experimental breakthroughs, have significantly expanded our understanding of information processing at the quantum level, leading to quantum cryptography \cite{ekert1992}, 
quantum teleportation \cite{bennett1993}, quantum metrology \cite{giovannetti2011}, quantum control \cite{guha2023}, and quantum computing \cite{raussendorf2001}. While fractional calculus offers a rich mathematical framework for describing complex phenomena, its application to quantum information problems remains relatively unexplored. In this sense, Zu and coworkers \cite{zu2021,zu2022} analyze the role of the memory effect of FTSE in the time evolution of a single quantum state and quantum entanglement by considering the Jaynes-Cummings (JC) model \cite{jaynes1963}, describing the interaction between a single atom and a single mode cavity. This interaction exhibits fascinating quantum phenomena like Rabi oscillations and entanglement, providing insights into fundamental light-matter interactions and laying the groundwork for advancements in quantum technologies. 
The two-level system interacting with the light field in the fractional scenario is also investigated in Refs. \cite{lu2017,lu2018}. 

Our work proposes a unitary associated with the fractional-time evolution of the JC model, guided by the formalism developed in \cite{cius22frac}. The key concepts underlying such formalism are revisited in Section \ref{sec:nonhermiticity}, where we briefly discuss the nonunitary nature of fractional time-evolution operator stemming from the Caputo derivative and establish a connection with TD non-Hermitian quantum formalism \cite{fring:16a,fring:17}. This connection is created by constructing a TD Dyson map, which is related to the dynamical Hilbert space metric. In Section \ref{sec:Unit}, we introduce the JC model and the fractional-time evolution operator by solving the FTSE and obtain an equivalent unitary time-evolution operator that describes the dynamics within the conventional quantum mechanical formulation. Subsequently, in Sections \ref{sec:population} and \ref{sec:entanglement}, we respectively analyze the collapse and revival phenomena, and quantify the entanglement dynamics of an atom interacting with a coherent field through the von Neumann entropy. Our conclusions follow in Section \ref{sec:Conc}.


\section{Non-hermiticity and the fractional-time scenario}
\label{sec:nonhermiticity}

The FTSE may generate many undesired results, such as the non-existence of stationary energy levels, non-unitarity of the evolution, and consequently, the non-conservation of probability, as discussed in Ref. \cite{laskin2017time}. It becomes evident when we transform the FTSE in a usual Schrödinger-like equation with an effective TD non-Hermitian Hamiltonian operator. It can be explicitly shown by applying the Riemann-Liouville derivative operator ${}^{\text{RL}}_{~~0}\mathcal{D}_t^{1-\alpha}$ on both sides of the Eq. \eqref{FTSE}, and utilizing the fractional differentiation property for $\alpha\in(0,1]$ \cite{iomin2019app}, expressed as:
\begin{equation}
{}^{\text{RL}}_{~~0}\mathcal{D}_t^{1-\alpha}\;^\text{C} _0\mathcal{D}_t^{\alpha}   \vert\Psi_{\alpha}(t)\rangle
= \partial_{t}\vert\Psi_{\alpha}(t)\rangle,
\end{equation}
where
\begin{equation}
{}^{\text{RL}}_{~~0}\mathcal{D}_t^{\alpha} (\cdot) = \frac{d}{dt} \int_{0}^{t}d\tau\frac{~(t-\tau)^{-\alpha}~}{\Gamma(1-\alpha)}(\cdot),
\end{equation}
Eq. \eqref{FTSE} becomes
\begin{align}
\label{TDSE0}
i\hbar \partial_{t}\vert\Psi_{\alpha}(t)\rangle
=
\frac{i^{1-\alpha}\hbar}{\hbar_{\alpha}}\hat{H}_{\alpha} {\,}^{\text{RL}}_{~~0}\mathcal{D}_t^{1-\alpha}\vert\Psi_{\alpha}(t)\rangle.
\end{align}
Note that the effective Hamiltonian is non-Hermitian, which implies a nonunitary time evolution of the quantum state. In this sense, different proposals have been made to map the fractional nonunitary evolution operator into a unitary one \cite{zhang2021quantization,laskin2017time,iomin2019fractional,cius22frac}. In particular, in Ref. \cite{cius22frac}, a unitary evolution for a traceless non-Hermitian two-level system evolving under FTSE was established by applying the TD non-Hermitian quantum formalism \cite{fring:16a,luiz:20,fring:17}. 

In the framework established in \cite{cius22frac}, a state undergoing nonunitary evolution, denoted by $\vert \Psi_{\alpha}(t) \rangle$, may be mapped to a state $\vert \psi_{\alpha}(t) \rangle$ evolving unitarily via the TD Dyson map $\hat{\eta}_{\alpha} (t)$. This transformation is expressed as follows
\begin{equation}
\label{psiPsiTDfrac}
\vert \psi_{\alpha} (t) \rangle 
=
\hat{\eta}_{\alpha} (t) \vert \Psi_{\alpha}(t) \rangle,
\end{equation}
such a map is assumed to be invertible. 
In what follows, from the fact that $\vert \psi_{\alpha} (t) \rangle = \hat{u}_{\alpha}(t)\vert \psi_{\alpha} (0) \rangle$ and $\vert \Psi_{\alpha}(t)\rangle = \hat{U}_{\alpha}(t) \vert\Psi_{\alpha}(0)\rangle$, the Eq. \eqref{psiPsiTDfrac} allows us to obtain the unitary time-evolution operator $\hat{u}_{\alpha}(t)$ in terms of the TD Dyson map and the nonunitary time-evolution operator as being
\begin{equation}
\label{uU}
\hat{u}_{\alpha}(t) = \hat{\eta}_{\alpha}(t)\hat{U}_{\alpha}(t)\hat{\eta}_{\alpha}^{-1}(0).
\end{equation}
Assuming the nonunitary time-evolution operator $\hat{U}_{\alpha}(t)$ is known, our primary goal is to identify the TD Dyson map parameters that enable the mapping of nonunitary dynamics to a unitary one. This can be accomplished by consider that $\hat{u}_{\alpha}(t)$ satisfies a Schrödinger-like equation
\begin{equation}
\label{u}
i\hbar\partial_{t}\hat{u}_{\alpha}(t)=\hat{h}_{\alpha}(t)\hat{u}_{\alpha}(t),
\end{equation}
and employing Eqs. \eqref{TDSE0}, \eqref{uU}, and \eqref{u} to obtain the Hermitian Hamiltonian $\hat{h}_{\alpha}(t)$, which is given by 
\begin{equation}
\label{hHer}
    \hat{h}_{\alpha}(t)
    =
    \hat{\eta}_{\alpha}(t)\hat{H}_{\alpha}^{\text{eff}}(t)\hat{\eta}_{\alpha}^{-1}(t)
    +
    i\hbar\partial_{t}\hat{\eta}_{\alpha}(t)\hat{\eta}_{\alpha}^{-1}(t),
\end{equation}
being $\hat{H}_{\alpha}^{\text{eff}}(t)$ the effective Hamiltonian into Eq. \eqref{TDSE0}:
\begin{equation}
    \hat{H}_{\alpha}^{\text{eff}}(t)
    =
    \frac{i^{1-\alpha}\hbar}{\hbar_{\alpha}}\hat{H}_{\alpha} {\,}^{\text{RL}}_{~~0}\mathcal{D}_t^{1-\alpha}\hat{U}_{\alpha}(t).
\end{equation}
Note that the effective Hamiltonian is not a pseudo-Hermitian operator and, therefore, it cannot be an observable \cite{fring:16a}. However, if a  TD Dyson map can be found that transforms the non-Hermitian Hamiltonian into a Hermitian one, it is possible to construct a dynamical Hilbert space equipped with a modified inner product, defined as $\langle \Psi_{\alpha}(t)\vert \Psi_{\alpha}(t)\rangle_{\Theta_{\alpha} (t)} = \langle \Psi_{\alpha}(t)\vert \hat{\Theta}_{\alpha} (t) \vert \Psi_{\alpha}(t)\rangle$, where the fractional-time evolution can be seen as a unitary in according to modified inner product
	\begin{align*}
	\langle \Psi_{\alpha}(t)\vert \Psi_{\alpha}(t)\rangle_{\Theta_{\alpha} (t)}
	&=
	\langle \Psi_{\alpha}(0)\vert \Psi_{\alpha}(0)\rangle_{\Theta_{\alpha} (0)}
	\\
	&=
	\langle \psi_{\alpha}(0) \vert \psi_{\alpha}(0)\rangle
	\\
	&=
	\langle \psi_{\alpha}(t) \vert \psi_{\alpha}(t)\rangle,
	\end{align*}
in which  $\hat{\Theta}_{\alpha}(t)= \hat{\eta}_{\alpha}^{\dagger}(t)\hat{\eta}_{\alpha}^{\,}(t)$ is the positive metric operator. 
The above relation reflects that the probability conservation in the fractional-time scenario can be achieved by defining a suitable TD metric with respect to which the state evolves unitarily. Moreover, this relation means that it is equivalent to mapping the state that evolves nonunitarily in another system that evolves unitarily to the usual metric. 
For more details about TD non-Hermitian systems see Refs. \cite{fring:16a,fring16b,fring:19,fring:17,luiz:20,mana:20,koussa:20,cius:22,cius2023}.

\section{Fractional-time dynamics of the Jaynes-Cummings model}
\label{sec:Unit}

\subsection{The model}
\label{subsec:Appl}
Now we analyze the JC model in the fractional-time scenario revised in the previous section. The JC model is a paradigmatic framework for understanding quantum light-matter interactions \cite{jaynes1963}. It describes the interaction of a two-level atom with a single quantized mode of the radiation field. In an ideal cavity QED experiment, the atom can be viewed as a two-level system ($\vert g \rangle$ and $\vert e \rangle$) interacting with a single mode of the field. This interaction is described by the JC Hamiltonian. Consider the resonant case, in which the energy difference between the two atomic levels exactly matches the energy of the radiation field. Within the rotating-wave approximation (RWA), the JC Hamiltonian in the interaction picture is given by
\begin{equation}
\label{H0}
\hat{H}_{\alpha}
=
\hbar_{\alpha}\mu_{\alpha}(\hat{\sigma}_{+}\hat{a} + \hat{\sigma}_{-}\hat{a}^{\dagger}),
\end{equation}
where the field is characterized by the annihilation $\hat{a}$ and creation $\hat{a}^\dagger$ bosonic operators satisfying the Weyl-Heisenberg algebra $[\hat{a},\hat{a}^\dagger]=1$. The operators $\hat{\sigma}_{+}=\vert e \rangle\langle g \vert$ and $\hat{\sigma}_{-}=\vert g \rangle\langle e \vert$ are the so-called atomic transition operators, which together with the inversion operator $\hat{\sigma}_{z}= \vert e \rangle\langle e \vert - \vert g \rangle\langle g \vert$ satisfy the $\mathfrak{su}(2)$ Lie algebra $[\hat{\sigma}_{+},\hat{\sigma}_{-}]=\hat{\sigma}_{z}$ and $[\hat{\sigma}_{z},\hat{\sigma}_{\pm}]=\pm2\hat{\sigma}_{\pm}$. The constant $\mu_\alpha$ denotes the atom-field coupling coefficient which
represents the strength of the atom-field coupling. 

The model is specified via the states of both atom and field, where the basis states of the field are the number states $\vert n \rangle$, with $n=0,1,2,\cdots$. In this case, the bare states $\vert g,n\rangle$ and $\vert e,n \rangle$ provide a natural basis for the infinite-dimensional Hilbert space representing the atom-field interaction. The ground state corresponds to the state with the atom in the ground state $\vert g \rangle$ and no photons in the cavity $\vert 0 \rangle$. In this case, we have the relation
\[
\hat{H}_{\alpha}\vert g,0\rangle=0,
\]
which means that spontaneous absorption from the vacuum is forbidden. Furthermore, for each photon number $n$, the Hamiltonian couples the bare states pairs $\vert e, n\rangle$ and $\vert g, n + 1 \rangle$, since
\begin{align*}
\hat{H}_{\alpha} \vert e,n\rangle
&=
\hbar_{\alpha}\mu_{\alpha}\sqrt{n+1}\vert g,n+1\rangle
\\
\hat{H}_{\alpha} \vert g,n+1\rangle
&=
\hbar_{\alpha}\mu_{\alpha}\sqrt{n+1}\vert e,n\rangle.
\end{align*}
Thus, the infinite-dimensional Hilbert space $\mathcal{H}$ consists of the one-dimensional subspace spanned by the ground state vector $\mathcal{H}_{\text{ground}}=\{\vert g, 0 \rangle\}$ and the mutually decoupled two-dimensional subspace $\mathcal{H}_{n}=\{\vert e, n\rangle, \vert g, n + 1 \rangle\}$. In other words, the Hilbert space $\mathcal{H} = L^2(\mathbb{R}) \otimes \mathbb{C}^2$ decays into a direct sum of dynamically invariant subspaces
\[
\mathcal{H}=\mathcal{H}_{\text{ground}}\oplus\mathcal{H}_{0}\oplus\mathcal{H}_{1}\oplus\mathcal{H}_{2}\oplus \cdots\,.
\] 

The Hamiltonian can be decomposed as a block-diagonal matrix
\begin{equation}
\label{H0_block}
\hat{H}_{\alpha} 
=
\hbar_{\alpha}
\left[
\begin{matrix}
0 & 0_{1\times2} & 0_{1\times2} & \cdots \\
0_{2\times1} & \hat{H}_{\alpha}^{(0)} & 0_{2\times2} & \cdots \\
0_{2\times1} & 0_{2\times2} & \hat{H}_{\alpha}^{(1)} &  \cdots \\
\vdots & \vdots & \vdots & \ddots \\
\end{matrix}
\right],
\end{equation}
where $\hat{H}_{\alpha}^{(n)}$ is the traceless $2\times 2$ matrix
\begin{equation}
\label{H0_matrix2d}
\hat{H}_{\alpha}^{(n)}
=
\hbar_{\alpha}\mu^{(n)}_{\alpha}\left[
\begin{matrix}
0 & 1 \\
1 & 0
\end{matrix}
\right],
\end{equation}
which represents the Hamiltonian of the system in the two-dimensional subspace $\mathcal{H}_{n}$, with $\mu^{(n)}_{\alpha} = \sqrt{n+1}\mu_\alpha$. A concise treatment of the JC model can be found in Ref. \cite{larson2021}.  Next, we discuss how the system evolves under the FTSE and can be described into the unitary framework.

\subsection{Fractional time-evolution}
\label{subsec:twolev}

In the fractional-time scenario, the dynamic of the system is claimed to be described by the FTSE given in Eq. \eqref{FTSE}. The formal solution of this equation can be read as $\vert \Psi_{\alpha}(t)\rangle =
\hat{U}_{\alpha}(t) \vert\Psi_{\alpha}(0)\rangle$, where the system evolves from an initial state $\vert\Psi^{\alpha}(0)\rangle$ to the state $\vert \Psi^{\alpha}(t)\rangle$ through the following time-evolution  operator $\hat{U}_{\alpha}(t)$,
\begin{eqnarray}
   \hat{U}_{\alpha}(t) = E_{\alpha}\left(i^{-\alpha}\hat{H}_{\alpha}\,t^{\alpha}/\hbar_{\alpha}\right),
\end{eqnarray}
that is a nonunitary operator satisfying the initial condition $\hat{U}_{\alpha}(0)=\hat{1}$. In the above equation, the function  $E_{\alpha}(x)= \sum_{k=0}^{\infty} x^{k}/\Gamma(\alpha k +1)$  is identified to be the well-known one-parameter Mittag-Leffler function \cite{podlubny1998fractional}. 

Once the Hamiltonian is represented in a block-diagonal form, as seen in Eq. \eqref{H0_block}, the nonunitary time-evolution operator can be represented as 
\begin{equation}
\label{U_block}
\hat{U}_{\alpha}(t) 
=
\left[
\begin{matrix}
1 & 0_{1\times2} & 0_{1\times2} & \cdots \\
0_{2\times1} & \hat{U}_{\alpha}^{(0)}(t) & 0_{2\times2} & \cdots \\
0_{2\times1} & 0_{2\times2} & \hat{U}_{\alpha}^{(1)}(t) & \cdots \\
\vdots & \vdots & \vdots & \ddots \\
\end{matrix}
\right],
\end{equation}
where $\hat{U}_{\alpha}^{(n)}(t)$ is given by
\begin{equation}
\label{U_matrix}
\hat{U}_{\alpha}^{(n)}(t)
=
\left[
\begin{matrix}
\mathcal{C}^{(n)}_{\alpha}(t) & i^{-\alpha}\mathcal{S}^{(n)}_{\alpha}(t) \\
i^{-\alpha}\mathcal{S}^{(n)}_{\alpha}(t) & \mathcal{C}^{(n)}_{\alpha}(t)
\end{matrix}
\right],
\end{equation}
satisfying the initial condition  $\hat{U}_{\alpha}^{(n)}(0)=\hat{1}_{2\times2}$. Here, the complex functions $\mathcal{C}^{(n)}_{\alpha}(t)$ and $\mathcal{S}^{(n)}_{\alpha}(t)$ are given in the form
\begin{subequations}
\begin{align}
    \mathcal{C}^{(n)}_{\alpha}(t)
  = {}
  &
    \frac{E_{\alpha}(i^{-\alpha}\mu^{(n)}_{\alpha}t^{\alpha}) + E_{\alpha}(-i^{-\alpha}\mu^{(n)}_{\alpha}t^{\alpha})}{2},
\\
    \quad
    \mathcal{S}^{(n)}_{\alpha}(t)
  = {}
  &
    \frac{E_{\alpha}(i^{-\alpha}\mu^{(n)}_{\alpha}t^{\alpha}) - E_{\alpha}(-i^{-\alpha}\mu^{(n)}_{\alpha}t^{\alpha})}{2i^{-\alpha}}.
\end{align}
\end{subequations}
While Naber’s approach to the FTSE is often regarded as non-physical, it remains a widely used framework for mathematical investigations. However, in the specific case of $\alpha=1/2$, the solution described by the Mittag-Leffler function acquires a meaningful physical interpretation, as discussed in Ref. \cite{iomin24}. This unique feature --where the $\alpha=1/2$ fractional order gains physical significance-- also arises in other approaches, as proposed in Refs. \cite{dias17,beims24}, where the fractional derivative emerges naturally within their respective frameworks.

The nonunitary nature of time evolution when considering the FTSE might be applied to mimic the effects of the environment on quantum systems. However, this treatment would be inherently heuristic, as nonunitarity leads to the non-conservation of probability. For a treatment of open quantum fractional dynamics, we suggest Ref. \cite{tarasov2010}, where a framework describing the nonunitary evolution of the density operator that is trace-preserving and completely positive for any initial condition is developed. Nevertheless, in this work, we are interested in establishing the conventional interpretation of quantum mechanics, by mapping the nonunitary fractional time-evolution operator to a unitary one by employing non-Hermitian quantum mechanics techniques with TD metrics \cite{cius22frac}. This procedure enables a proper quantum-mechanical interpretation of the fractional-time description, utilizing a modified inner product as done in the non-Hermitian framework \cite{fring:16a,fring:17}.

\subsection{Unitary evolution}

Indeed, the choice of TD Dyson map is not unique. To address this, we propose a Hermitian time-dependent Dyson map,  $\hat{\eta}_{\alpha}(t)$, in block-diagonal form, expressed as:
\begin{equation}
\label{Dyson_block}
\hat{\eta}_{\alpha}(t) 
=
\left[
\begin{matrix}
1 & 0_{1\times2} & 0_{1\times2} & \cdots \\
0_{2\times1} & \hat{\eta}_{\alpha}^{(0)}(t) & 0_{2\times2} & \cdots \\
0_{2\times1} & 0_{2\times2} & \hat{\eta}_{\alpha}^{(1)}(t) & \cdots \\
\vdots & \vdots & \vdots & \ddots \\
\end{matrix}
\right],
\end{equation}
with the TD Dyson map acting on the two-dimensional subspace $\mathcal{H}_{n}$ chosen as  
\begin{equation*}
\label{DysonGD}
  \hat{\eta}_{\alpha}^{(n)}(t) =
  e^{\kappa_{\alpha}^{(n)}(t)}
  e^{\lambda_{\alpha}^{(n)}(t)\hat{\sigma}_{+}}
  e^{\ln{\Lambda_{\alpha}^{(n)}(t)}\hat{\sigma}_{z}/2}
  e^{[\lambda_{\alpha}^{(n)}(t)]^{\ast}\hat{\sigma}_{-}},
\end{equation*}
where we assume $\lambda_{\alpha}^{(n)}(t) \in \mathbb{C}$ and $\kappa_{\alpha}^{(n)}(t), \Lambda_{\alpha}^{(n)}(t) \in \mathbb{R}$ with the additional condition $\Lambda_{\alpha}^{(n)}(t) > 0$. Furthermore, we can represent the $n$-th subspace of the time-dependent Dyson map as a $2 \times 2$ matrix in the basis $\{\vert e, n\rangle , \vert g, n + 1 \rangle\}$, yielding:
\begin{equation}
\label{etamat}
\hat{\eta}_{\alpha}^{(n)}(t) =        \frac{e^{\kappa_{\alpha}^{(n)}(t)}}{\sqrt{\Lambda_{\alpha}^{(n)}(t)}}\left[
        \begin{matrix}
        \chi_{\alpha}^{n}(t) & \lambda_{\alpha}^{(n)}(t) \\
        [\lambda_{\alpha}^{(n)}(t)]^{\ast} & 1
\end{matrix} \right]\rev{,}
\end{equation}
being the function $\chi_{\alpha}^{(n)}(t)=\Lambda_{\alpha}^{(n)}(t) + |\lambda_{\alpha}^{(n)}(t)|^{2}$.

By applying the results of Eqs. \eqref{U_matrix} and \eqref{etamat} in Eq. \eqref{uU}, we represent the unitary operator $\hat{u}_{\alpha}(t)$ as
\begin{equation}
\label{u_block}
\hat{u}_{\alpha}(t) 
=
\left[
\begin{matrix}
1 & 0_{1\times2} & 0_{1\times2} & \cdots \\
0_{2\times1} & \hat{u}_{\alpha}^{(0)}(t) & 0_{2\times2} & \cdots \\
0_{2\times1} & 0_{2\times2} & \hat{u}_{\alpha}^{(1)}(t) & \cdots \\
\vdots & \vdots & \vdots & \ddots \\
\end{matrix}
\right],
\end{equation}
with $\hat{u}_{\alpha}^{(n)}(t) = \hat{\eta}_{\alpha}^{(n)}(t)\hat{U}_{\alpha}^{(n)}(t)[\hat{\eta}_{\alpha}^{(n)}(0)]^{-1}$. Admitting $\hat{u}_{\alpha}(t)$ as a unitary operator, it implies that $\hat{u}_{\alpha}^{(n)}(t)$ must necessarily belong to the Lie group $U(2)$ and can be represented in the general matrix form
\begin{equation}
\label{u_matrix}
\hat{u}^{(n)}_{\alpha}(t)
=
e^{i\delta_{\alpha}^{(n)}(t)}
\left[
\begin{matrix}
\varpi_{\alpha,+}^{(n)}(t) & \varpi_{\alpha,-}^{(n)}(t) \\
-[\varpi_{\alpha,-}^{(n)}(t)]^{\ast} & [\varpi_{\alpha,+}^{(n)}(t)]^{\ast}
\end{matrix}
\right],
\end{equation}
with 
\begin{subequations}
	\label{u_coeff}
	\begin{align}
	\label{u_coeff1}
	\delta_{\alpha}^{(n)}(t) = {}
	&
	\frac{1}{2}\mathrm{Im}         [\ln{D_{\alpha}^{(n)}(t)}],
        \\
	\label{u_coeff2}
        \varpi_{\alpha,\pm}^{(n)}(t) = {}
	&
	\pm e^{i\delta_{\alpha}^{(n)}(t)}[\nu_{\alpha,\mp}^{(n)}(t)]^{\ast} ,
	\end{align}
\end{subequations}
where the function $D_{\alpha}^{(n)}(t)$ in Eq. \eqref{u_coeff1} is defined as $D_{\alpha}^{(n)}(t) = 
[\mathcal{C}^{(n)}_{\alpha}(t)]^{2}
-
(-1)^{-\alpha}[\mathcal{S}^{(n)}_{\alpha}(t)]^{2}$.
To simplify the notation, from now on we omit the time dependence and define $\nu_{\alpha,\pm}^{(n)}$ in Eq. \eqref{u_coeff2} as follows:
  \begin{align}
  \label{coeu}
    \nu_{\alpha,\pm}^{(n)}  = {}
    &
      \pm\frac{e^{\kappa_{\alpha}^{(n)}-\kappa_{\alpha}^{(n)}(0)}}{\sqrt{\Lambda_{\alpha}^{(n)}\Lambda_{\alpha}^{(n)}(0)}}
        \left[
        \zeta_{\alpha,\pm}^{(n)} + 
        (\lambda_{\alpha}^{(n)})^{\ast} \xi_{\alpha,\pm}^{(n)}
        \right],
        \end{align}
where $\zeta_{\alpha,\pm}^{(n)}$ and $\xi_{\alpha,\pm}^{(n)}$ are given by
\begin{subequations}
        \begin{align}
        \zeta_{\alpha,+}^{(n)} &=  i^{-\alpha}\mathcal{S}_{\alpha}^{(n)} - [\lambda_{\alpha}^{(n)}(0)]^{\ast} \mathcal{C}_{\alpha}^{(n)},
        \\
        \zeta_{\alpha,-}^{(n)} &= i^{-\alpha}\lambda_{\alpha}^{(n)}(0) \mathcal{S}_{\alpha}^{(n)} - \chi_{\alpha}^{(n)}(0) \mathcal{C}_{\alpha}^{(n)},
        \\
        \xi_{\alpha,+}^{(n)} &=  \mathcal{C}_{\alpha}^{(n)} - i^{-\alpha}[\lambda_{\alpha}^{(n)}(0)]^{\ast} \mathcal{S}_{\alpha}^{(n)} ,
        \\
        \xi_{\alpha,-}^{(n)} &= \lambda_{\alpha}^{(n)}(0) \mathcal{C}_{\alpha}^{(n)} -  i^{-\alpha}\chi_{\alpha}^{(n)}(0) \mathcal{S}_{\alpha}^{(n)},
        \end{align}
\end{subequations}
Moreover, for $\hat{u}_{\alpha}^{(n)}$ to represent a unitary operator, the parameters of the TD Dyson map must take the form:
\begin{subequations}
    \begin{align}
	\kappa_{\alpha}^{(n)} = {}
	&
	\kappa_{\alpha}^{(n)}(0) -\frac{1}{2}\mathrm{Re}[\ln{D_{\alpha}^{(n)}}],
	\\
	\chi_{\alpha}^{(n)} = {}
	&
        \frac{|\zeta_{\alpha,+}^{(n)}|^{2} + |\zeta_{\alpha,-}^{(n)}|^{2} + \Lambda_{\alpha}^{(n)}(0)e^{\mathrm{Re}[\ln{D_{\alpha}^{(n)}}]}}{ |\xi_{\alpha,+}^{(n)}|^{2} + |\xi_{\alpha,-}^{(n)}|^{2} + \Lambda_{\alpha}^{(n)}(0)e^{\mathrm{Re}[\ln{D_{\alpha}^{(n)}}]}},
	\\
	\lambda_{\alpha}^{(n)} = {}
	&
	\frac{-\left[\xi_{\alpha,+}^{(n)}[\zeta_{\alpha,+}^{(n)}]^{\ast} +
		\xi_{\alpha,-}^{(n)}[\zeta_{\alpha,-}^{(n)}]^{\ast}\right]}{|\xi_{\alpha,+}^{(n)}|^{2}
		+
		|\xi_{\alpha,-}^{(n)}|^{2}
		+ \Lambda_{\alpha}^{(n)}(0)e^{\mathrm{Re}[\ln{D_{\alpha}^{(n)}}]}},
    \end{align}
\end{subequations}
where the TD Dyson map parameter $\Lambda_{\alpha}^{(n)} =  \chi_{\alpha}^{(n)}-|\lambda_{\alpha}^{(n)}|^{2}$, with the condition that $\chi_{\alpha}^{(n)}>|\lambda_{\alpha}^{(n)}|^{2}$, since we are assuming $\Lambda_{\alpha}^{(n)}$ as a positive function. The functions in Eq. \eqref{u_coeff2} satisfy the relation
$
|\varpi_{\alpha,+}^{(n)}|^{2} + |\varpi_{\alpha,-}^{(n)}|^{2} = 1
$, 
which comes from the fact of $|\det\hat{u}^{(n)}_{\alpha}|^{2}=1$. Note that the coefficients of the unitary operator $\hat{u}^{(n)}_{\alpha}$ depend only on the parameters of the fractional-time evolution operator and the initial values of the parameters of the TD Dyson map, which are incorporated into the functions $\xi_{\alpha,\pm}^{(n)}$ and $\zeta_{\alpha,\pm}^{(n)}$. For more details on those calculations, see Ref. \cite{cius22frac}.

\section{Population Inversion}
\label{sec:population}

\begin{figure*}[ht!]
    \begin{center}
    \includegraphics[width=0.48\linewidth]
 {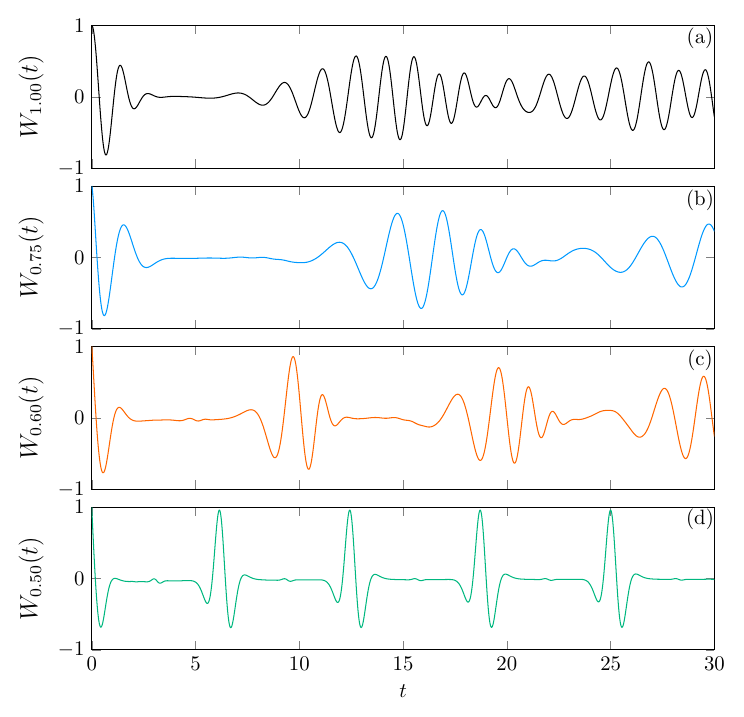}
 \includegraphics[width=0.48\linewidth]
 {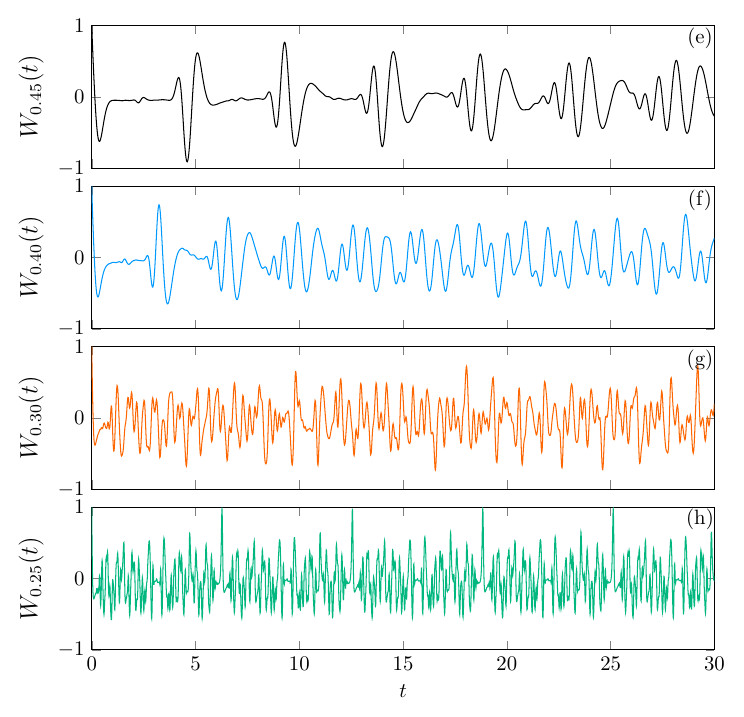}
    \end{center}
    \caption{
    \label{fig:CollapseRevivals}
    Time evolution of the population inversion $W_{\alpha}(t)$ for different values of $\alpha$. It is considered initially the atom in the excited state and the field in the coherent state with $\beta=2$, and the initial values of Dyson map parameters being $\kappa_{\alpha}(0)=0$, $\Lambda_{\alpha}(0)=1$ and $\lambda_{\alpha}(0)=0$. The atom-field coupling is $\mu_{\alpha}=1$. Then, we analyze for $\alpha=1.00$ (a), $\alpha=0.75$ (b), $\alpha=0.60$ (c), and $\alpha=0.50$ (d), $\alpha=0.45$ (e), $\alpha=0.40$ (f), $\alpha=0.30$ (g), and $\alpha=0.25$ (h). 
    }
\end{figure*}

The collapse and revival of atomic oscillations is a distinctive feature observed in the interaction of a two-level atom with a quantized electromagnetic field inside a cavity described by the JC model \cite{larson2021}. 
When the cavity field is initially prepared in a coherent state and interacts with the two-level atom, the system exhibits a fascinating dynamical behavior characterized by the following stages:
i) initial Rabi oscillations: at the beginning, the atom exchanges energy with the quantized field, resulting in oscillations of the atomic population between the ground and excited states;
ii) collapse: the distribution of photon number states within the coherent field leads to dephasing, causing the Rabi oscillations to decay; iii) revival: after a certain period, the oscillations rephase, and the atomic population oscillations reappear. 
The timing of these revivals is determined by the properties of the coherent state and the parameters of the system.

We aim to analyze the atomic population inversion depending on the $\alpha$ parameter within a unitary framework. We consider an initial state where the atom is excited and the field is in a coherent state: $\vert \psi_{\alpha}(0) \rangle = \vert e \rangle\otimes \vert \beta \rangle$, where $\vert \beta \rangle=\sum_{n=0}^{\infty}c_n\vert n \rangle$ where $c_n=e^{-|\beta|^2/2}\beta^{n}/\sqrt{n!}$ with $\beta\in\mathbb{C}$. The system evolves in according to $\vert \psi_{\alpha}(t) \rangle = \hat{u}_{\alpha}(t)\vert \psi_{\alpha}(0) \rangle$, leading to
\begin{equation}
\label{state_t}
\vert \psi_{\alpha}(t) \rangle = \sum_{n=0}^{\infty}  \Big[A_{e,n}^{\alpha}(t)\vert e , n\rangle + A_{g,n}^{\alpha}(t)\vert g , n + 1\rangle\Big],
\end{equation}
where the probability amplitudes are
\begin{subequations}
	\begin{align}
		A_{e,n}^{\alpha}(t)&=c_{n}e^{i\delta_{\alpha}^{(n)}(t)}\varpi_{\alpha,+}^{(n)}(t), 
		\\
		A_{g,n}^{\alpha}(t)&=-c_{n}e^{i\delta_{\alpha}^{(n)}(t)}[\varpi_{\alpha,-}^{(n)}(t)]^{\ast}.
	\end{align}
\end{subequations}
The probability of finding the atom in the excited state with the field having $n$ photons is $P_{e,n}^{\alpha}(t)=|A_{e,n}^{\alpha}(t)|^2$. In contrast, the probability of finding the atom in the ground state with the field having $n+1$ photons is $P_{g,n+1}^{\alpha}(t)=|A_{g,n}^{\alpha}(t)|^2$. Furthermore, we can marginalize the probability over the field states, by summing over all possible photon numbers to obtain the probability of finding the atom in an excited or grounded state, which are given respectively by $P_{e}^{\alpha}(t)=\sum_{n=0}^{\infty}|A_{e,n}^{\alpha}(t)|^2$ and $P_{g}^{\alpha}(t)=\sum_{n=0}^{\infty}|A_{g,n}^{\alpha}(t)|^2$ .
We then calculate the population inversion over time, which is given by the mean value of the inversion operator, 
\begin{align}
\label{PopInv}
W_{\alpha} (t) 
&= P_{e}^{\alpha}(t)-P_{g}^{\alpha}(t)
\nonumber\\
&=\sum_{n=0}^{\infty}\left[|A_{e,n}^{\alpha}(t)|^2 - |A_{g,n}^{\alpha}(t)|^2\right]. 
\end{align}

In Fig. \ref{fig:CollapseRevivals}, we illustrate the time evolution of the population inversion $W_{\alpha} (t)$ by considering the atom-field coupling $\mu_{\alpha}=1$ for a coherent state with $\beta=2$ and different values of the parameter $\alpha$. The initial conditions for the TD Dyson map are set as $\kappa_{\alpha}^{(n)}(0)=0$, $\Lambda_{\alpha}^{(n)}(0)=1$, and $\lambda_{\alpha}^{(n)}(0)=0$. 
These conditions imply that the Dyson map at $t=0$,  $\hat{\eta}_{\alpha}(0)$, is the identity operator. In this case, the initial states in both the unitary and nonunitary representations are identical, thus, $\vert\psi_{\alpha}(0)\rangle = \vert\Psi_{\alpha}(0)\rangle$. 
As $\alpha$ decreases, the dynamics begin to change. 
For instance, when $\alpha=1.00$ (Fig. \ref{fig:CollapseRevivals}a), we recover the well-known population inversion dynamics of the JC model, as the Caputo derivative reduces to the first-order derivative in Eq. \eqref{FTSE}. For $\alpha=0.75$ (Fig. \ref{fig:CollapseRevivals}b), the population inversion exhibits a slightly modified behavior with a longer collapse time and fewer oscillations. For $\alpha=0.60$ (Fig. \ref{fig:CollapseRevivals}c),  irregular collapse and revival patterns emerge. Interestingly, when $\alpha=0.50$ (Fig. \ref{fig:CollapseRevivals}d), the population inversion exhibits a more periodic behavior, returning close to the excited state with a period of approximately $6.18$. One might expect that further decreasing $\alpha$ would lead to even more regular behavior. However, for $\alpha<0.5$ the dynamics become highly irregular. Sudden peaks or troughs characterize the population inversion, and the characteristic collapse and revival patterns diminish due to the large fluctuations. For instance, when $\alpha=0.45$ (Fig. \ref{fig:CollapseRevivals}e), the regular pattern observed for $\alpha=0.50$ disappears. A near-complete population inversion occurs after the initial collapse, followed by fewer subsequent collapses. For $\alpha=0.40$ (Fig. \ref{fig:CollapseRevivals}f) and $\alpha=0.30$ (Fig. \ref{fig:CollapseRevivals}f) an irregular oscillations pattern with faster oscillations manifests. When $\alpha=0.25$, the population inversion exhibits a pattern of faster oscillations, but the overall dynamics remain quasi-periodic.

Remarkably, within the unitary framework, the unique properties of fractional derivatives can be attributed to additional driving forces in the atom-field interaction \cite{dutra1993}. These forces arise from combining the fractional-derivative parameter and the TD Dyson map. This can be inferred by representing the system in the $\mathcal{H}_{n}$ basis, where the Hermitian Hamiltonian in Eq. \eqref{hHer} takes the form $\hat{h}_{\alpha}^{(n)}(t) \propto \mu_{i,\alpha}^{(n)}(t)\hat{\sigma}_{i}$ with $i=x,y,z$.

\section{Atom-field Entanglement}
\label{sec:entanglement}

The entanglement phenomenon has been discussed since the beginning of quantum mechanics \cite{schrodinger1935}. It describes a scenario in which the quantum states of two or more particles become so correlated that the state of one particle cannot be described independently of the others, even when spatially separated. This is traditionally viewed as a manifestation of non-separability \cite{horodecki2009}. In Bell nonlocality, not all entangled states violate Bell's inequality, but any state that does violate it must be entangled. Thus, entanglement is a necessary condition for violating Bell’s inequality \cite{werner1989}. This violation reflects deviations in the statistical correlations of quantum states from classical expectations based on local realism \cite{einstein1935,bell1964}.  These concepts underscore the nonclassical nature of quantum mechanics and carry profound implications for our understanding of reality. 

We explore the influence of the FTSE on quantum entanglement. Previous studies \cite{zu2021, zu2022} have explored entanglement within the JC model using FTSE. Here, we analyze the entanglement dynamics based on the unitary framework associated with the FTSE, using the TD Dyson map as previously discussed. To quantify entanglement in pure bipartite states, we employ the von Neumann entropy \cite{petz2001}, a well-established measure of entanglement, defined as:
\begin{equation}
\label{entropy}
  S(\hat{\rho}_{i})=-\Tr(\hat{\rho}_{i}\ln\hat{\rho}_{i}),
\end{equation}
where $\hat{\rho}_{i}$ represents the reduced state of the subsystem $i$ (with $i=\text{a, f}$, referring to the atom and field, respectively). For separable states, the von Neumann entropy yields a zero value, indicating the absence of entanglement. In contrast, for entangled states, the entropy returns a positive value, signifying the presence of nonclassical correlations within the system. In this case, the von Neumann entropy is symmetric with respect to the partitions, leading to   $S(\hat{\rho}_{\text{a}})=S(\hat{\rho}_{\text{f}})$. 

The composite state given in Eq. \eqref{state_t} corresponds to the density operator
\begin{equation}
\hat{\rho}^{\alpha}(t) =\vert \psi_{\alpha}(t) \rangle \langle \psi_{\alpha}(t) \vert.
\end{equation}
Performing a partial trace over the field subsystem, we obtain the reduced density matrix of the atom, $\hat{\rho}_{\text{a}}^{\alpha}(t)$, given by
\begin{equation}
\label{density_reduced}
\hat{\rho}_{\text{a}}^{\alpha}(t) =
\left[
\begin{matrix}
P_{e}^{\alpha}(t) & r_{\alpha}(t) \\
[r_{\alpha}(t)]^{\ast} & P_{g}^{\alpha}(t)
\end{matrix}
\right],
\end{equation}
where the diagonal terms, $P_{e}^{\alpha}(t)$ and $P_{g}^{\alpha}(t)$, represent the population of the excited and grounded states, respectively, as discussed earlier. The off-diagonal coherence term is given by $r_{\alpha}(t) = \sum_{n=0}^{\infty}A_{e,n+1}^{\alpha}[A_{g,n}^{\alpha}]^{\ast}$. Note that the trace-preserving property of the reduced density matrix is verified, as $P_{e}^{\alpha}(t) + P_{g}^{\alpha}(t) = 1$. 

\begin{figure*}[t!]
    \begin{center}
	\includegraphics[width=0.48\linewidth]
 {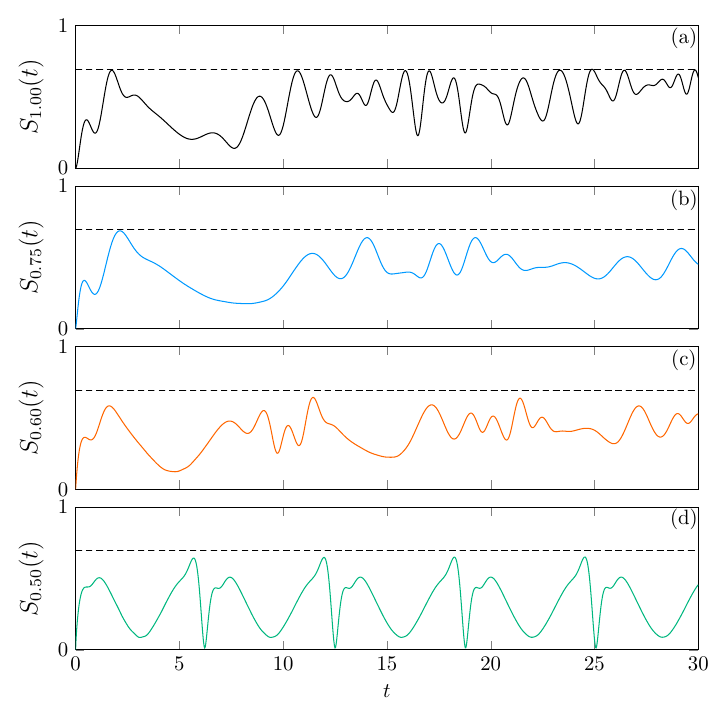}
 \includegraphics[width=0.48\linewidth]
 {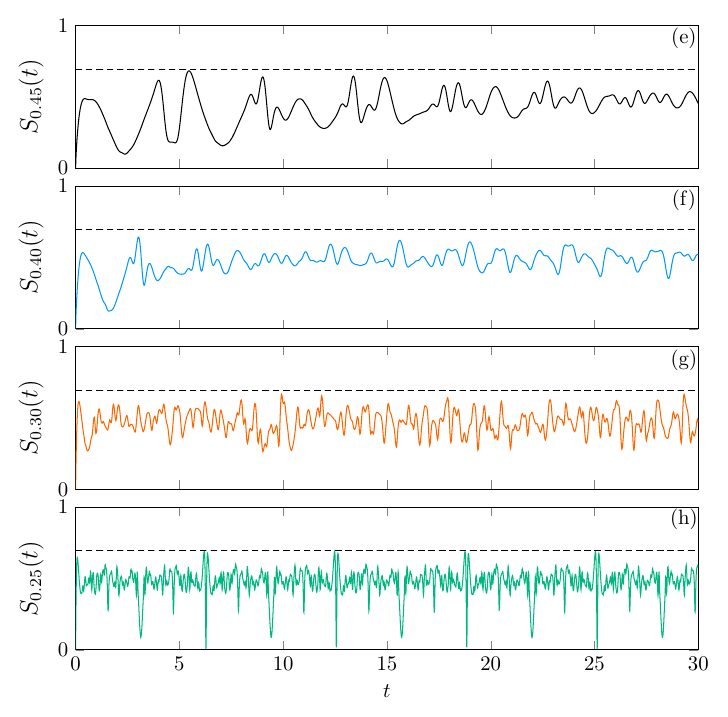}
    \end{center}
    \caption{
    \label{fig:Entropy}
    Time evolution of the von Neumann entropy $S_{\alpha}(t)$ for different values of $\alpha$. The horizontal dashed line indicates the upper bound for the von Neumann entropy $S_{\text{max}}=\ln2\approx0.693$. It is considered initially the atom in the excited state and the field in the coherent state with $\beta=2$, and the initial values of Dyson map parameters being $\kappa_{\alpha}(0)=0$, $\Lambda_{\alpha}(0)=1$ and $\lambda_{\alpha}(0)=0$, with the atom-field coupling $\mu_{\alpha}=1$. We analyze for $\alpha=1.00$ (a), $\alpha=0.75$ (b), $\alpha=0.60$ (c), and $\alpha=0.50$ (d), $\alpha=0.45$ (e), $\alpha=0.40$ (f), $\alpha=0.30$ (g), and $\alpha=0.25$ (h). }
\end{figure*}

For the density matrix in Eq. \eqref{density_reduced}, we denote the von Neumann entropy by $S_{\alpha}(t)$. Then, from Eq. \eqref{entropy}, it follows that
\begin{equation}
\label{vNentropy}
    S_{\alpha}(t) = -\vartheta_{+}^{\alpha}(t)\ln\vartheta_{+}^{\alpha}(t) - \vartheta_{-}^{\alpha}(t)\ln\vartheta_{-}^{\alpha}(t),
\end{equation}
where the $\vartheta_{\pm}^{\alpha}(t)$ denote the eigenvalues of the reduced density matrix  $\hat{\rho}_{\text{a}}^{\alpha}(t)$, given by
\begin{equation}
    \vartheta_{\pm}(t) = \frac{1}{2} \pm \frac{1}{2}\sqrt{W_{\alpha}^{2}(t)+4|r_{\alpha}(t)|^{2}},
\end{equation}
with $W_{\alpha}(t)$ being the population inversion expressed in Eq. \eqref{PopInv}. To quantify the entanglement dynamics, we numerically evaluate the von Neumann entropy, as defined in Eq. \eqref{vNentropy}, for different values of $\alpha$. The results are depicted in Fig. \ref{fig:Entropy}. In the plot, the horizontal dashed line indicates the upper bound for the von Neumann entropy for a two-level system, which is given by $S_{\text{max}}=\ln2\approx0.693$. Similar to the previous analysis, we set the atom-field coupling to be  $\mu_{\alpha}=1$, the coherent parameter $\beta=2$, and the initial TD Dyson map parameters to be $\kappa_{\alpha}(0)=0$, $\Lambda_{\alpha}(0)=1$ and $\lambda_{\alpha}(0)=0$. 
The case of $\alpha = 1.00$ (Fig. \ref{fig:Entropy}a) corresponds to the well-known JC model. Starting with a separable pure state for the field and atom, the quantum dynamics, for $t > 0$, lead to increasing marginal entropy and strong entanglement between the field and the atom \cite{phoenix1988,phoenix1991s}. In this case, the entanglement reaches its maximum value at various instants of time, while for $1>\alpha\geq0$, it approaches the maximum with a slight deviation. For $\alpha = 0.75$ (Fig. \ref{fig:Entropy}b) the entanglement reaches its maximum value only at the beginning of evolution and remains oscillating below the upper bound. When $\alpha = 0.60$ (Fig. \ref{fig:Entropy}c), the entanglement oscillates without reaching its maximum value within the plotted time. For $\alpha = 0.50$ (Fig. \ref{fig:Entropy}d), the entanglement exhibits a more periodic pattern. It is characterized by an initial rise in entropy, subsequent oscillations, and a quick return to a nearly separable state, suggesting entanglement birth and death cycles. As observed in the case of the population inversion, for $\alpha<0.50$, large oscillations start to appear in entanglement entropy. This is illustrated for $\alpha=0.45$ (Fig. \ref{fig:Entropy}e), $\alpha=0.40$ (Fig. \ref{fig:Entropy}e), $\alpha=0.30$ (Fig. \ref{fig:Entropy}g), and $\alpha=0.25$ (Fig. \ref{fig:Entropy}h). Despite large fluctuations, the behavior for $\alpha=0.25$ is quasi-periodic.


\section{Conclusions}
\label{sec:Conc}
To summarize, we describe the dynamics of the JC model within a fractional-time to a unitary framework, employing the Caputo derivative as outlined in Ref. \cite{cius22frac} for a traceless two-level system with a general non-Hermitian Hamiltonian. This framework links the FTSE to the TD non-Hermitian Hamiltonian theory \cite{fring:16a,fring:17}, where a TD Dyson map is constructed to define a dynamic Hilbert space with a modified inner product with respect to which the fractional time evolution is unitary. It circumvents the issue of nonunitarity and aligns with quantum mechanical postulates. To apply the results from Ref. \cite{cius22frac} to the infinite-dimensional JC model, we decompose it into dynamical invariant two-dimensional subspaces. We then examine how the fractional-order parameters within the unitary dynamics affect the collapse and revival phenomenon, which are intrinsic to the quantum nature of the electromagnetic field and absent in classical systems. We also explore the atom-field entanglement across different fractional-order parameters, revealing the potential for both the birth and death of entanglement. Entanglement, a genuine quantum phenomenon with no classical analog, provides valuable insights into quantum mechanics and is of significant interest in quantum optics and quantum information science. 
As a potential application, the population inversion dynamics described by Eq. \eqref{PopInv} could be emulated using a reverse engineering approach, as presented in Ref. \cite{sun2006}. This approach involves designing a time-dependent control field within the standard JC model to achieve a specific target population inversion profile. 
As a further idea, this treatment presented here also may be applied to explore the $\kappa$-deformed JC model, where the JC interaction is described by a time-independent non-Hermitian Hamiltonian due to the deformation parameter \cite{uhdre2022}. We hope our analysis stimulates further research on unitary formulations for fractional-time and non-Hermitian systems, potentially leading to new insights into fundamental physics.

\section*{Acknowledgements}
The author thanks Maike A. F. dos Santos, Antonio S. M. de Castro, and  Thiago T. Tsutsui for their valuable comments and clarifications that significantly improved the work. Thanks are also due to Bárbara Amaral and the members of the Quantum Foundation Group (IFUSP) for their insightful discussions. Additionally, the author would like to acknowledge financial support from Instituto Serrapilheira, and the Pró-Reitoria de Pesquisa e Inovação (PRPI) from the Universidade de São Paulo (USP) by financial support through the Programa de Estímulo à Supervisão de Pós-Doutorandos por Jovens Pesquisadores.



\bibliographystyle{apsrev4-2}
\bibliography{references.bib}

\begin{thebibliography}{61}%
\makeatletter
\providecommand \@ifxundefined [1]{%
 \@ifx{#1\undefined}
}%
\providecommand \@ifnum [1]{%
 \ifnum #1\expandafter \@firstoftwo
 \else \expandafter \@secondoftwo
 \fi
}%
\providecommand \@ifx [1]{%
 \ifx #1\expandafter \@firstoftwo
 \else \expandafter \@secondoftwo
 \fi
}%
\providecommand \natexlab [1]{#1}%
\providecommand \enquote  [1]{``#1''}%
\providecommand \bibnamefont  [1]{#1}%
\providecommand \bibfnamefont [1]{#1}%
\providecommand \citenamefont [1]{#1}%
\providecommand \href@noop [0]{\@secondoftwo}%
\providecommand \href [0]{\begingroup \@sanitize@url \@href}%
\providecommand \@href[1]{\@@startlink{#1}\@@href}%
\providecommand \@@href[1]{\endgroup#1\@@endlink}%
\providecommand \@sanitize@url [0]{\catcode `\\12\catcode `\$12\catcode
  `\&12\catcode `\#12\catcode `\^12\catcode `\_12\catcode `\%12\relax}%
\providecommand \@@startlink[1]{}%
\providecommand \@@endlink[0]{}%
\providecommand \url  [0]{\begingroup\@sanitize@url \@url }%
\providecommand \@url [1]{\endgroup\@href {#1}{\urlprefix }}%
\providecommand \urlprefix  [0]{URL }%
\providecommand \Eprint [0]{\href }%
\providecommand \doibase [0]{https://doi.org/}%
\providecommand \selectlanguage [0]{\@gobble}%
\providecommand \bibinfo  [0]{\@secondoftwo}%
\providecommand \bibfield  [0]{\@secondoftwo}%
\providecommand \translation [1]{[#1]}%
\providecommand \BibitemOpen [0]{}%
\providecommand \bibitemStop [0]{}%
\providecommand \bibitemNoStop [0]{.\EOS\space}%
\providecommand \EOS [0]{\spacefactor3000\relax}%
\providecommand \BibitemShut  [1]{\csname bibitem#1\endcsname}%
\let\auto@bib@innerbib\@empty
\bibitem [{\citenamefont {Podlubny}(1998)}]{podlubny1998fractional}%
  \BibitemOpen
  \bibfield  {author} {\bibinfo {author} {\bibfnamefont {I.}~\bibnamefont
  {Podlubny}},\ }\href@noop {} {\emph {\bibinfo {title} {Fractional
  differential equations}}}\ (\bibinfo  {publisher} {Academic Press},\ \bibinfo
  {year} {1998})\BibitemShut {NoStop}%
\bibitem [{\citenamefont {Chaves}(1998)}]{chaves1998}%
  \BibitemOpen
  \bibfield  {author} {\bibinfo {author} {\bibfnamefont {A.}~\bibnamefont
  {Chaves}},\ }\href
  {https://doi.org/https://doi.org/10.1016/S0375-9601(97)00947-X} {\bibfield
  {journal} {\bibinfo  {journal} {Phys. Lett. A}\ }\textbf {\bibinfo {volume}
  {239}},\ \bibinfo {pages} {13} (\bibinfo {year} {1998})}\BibitemShut
  {NoStop}%
\bibitem [{\citenamefont {Metzler}\ and\ \citenamefont
  {Klafter}(2000)}]{metzler2000random}%
  \BibitemOpen
  \bibfield  {author} {\bibinfo {author} {\bibfnamefont {R.}~\bibnamefont
  {Metzler}}\ and\ \bibinfo {author} {\bibfnamefont {J.}~\bibnamefont
  {Klafter}},\ }\href
  {https://doi.org/https://doi.org/10.1016/S0370-1573(00)00070-3} {\bibfield
  {journal} {\bibinfo  {journal} {Phys. Rep.}\ }\textbf {\bibinfo {volume}
  {339}},\ \bibinfo {pages} {1} (\bibinfo {year} {2000})}\BibitemShut {NoStop}%
\bibitem [{\citenamefont {Laskin}(2007)}]{laskin2007}%
  \BibitemOpen
  \bibfield  {author} {\bibinfo {author} {\bibfnamefont {N.}~\bibnamefont
  {Laskin}},\ }\href
  {https://doi.org/https://doi.org/10.1016/j.cnsns.2006.01.001} {\bibfield
  {journal} {\bibinfo  {journal} {Commun. Nonlinear Sci. Numer. Simul.}\
  }\textbf {\bibinfo {volume} {12}},\ \bibinfo {pages} {2} (\bibinfo {year}
  {2007})}\BibitemShut {NoStop}%
\bibitem [{\citenamefont {Longhi}(2015)}]{longhi2015fractional}%
  \BibitemOpen
  \bibfield  {author} {\bibinfo {author} {\bibfnamefont {S.}~\bibnamefont
  {Longhi}},\ }\href {https://doi.org/https://doi.org/10.1364/OL.40.001117}
  {\bibfield  {journal} {\bibinfo  {journal} {Opt. Lett}\ }\textbf {\bibinfo
  {volume} {40}},\ \bibinfo {pages} {1117} (\bibinfo {year}
  {2015})}\BibitemShut {NoStop}%
\bibitem [{\citenamefont {Huang}\ and\ \citenamefont
  {Dong}(2017)}]{huang2017beam}%
  \BibitemOpen
  \bibfield  {author} {\bibinfo {author} {\bibfnamefont {C.}~\bibnamefont
  {Huang}}\ and\ \bibinfo {author} {\bibfnamefont {L.}~\bibnamefont {Dong}},\
  }\href {https://doi.org/https://doi.org/10.1038/s41598-017-05926-5}
  {\bibfield  {journal} {\bibinfo  {journal} {Sci. Rep.}\ }\textbf {\bibinfo
  {volume} {7}},\ \bibinfo {pages} {1} (\bibinfo {year} {2017})}\BibitemShut
  {NoStop}%
\bibitem [{\citenamefont {Zhang}\ \emph {et~al.}(2016)\citenamefont {Zhang},
  \citenamefont {Zhong}, \citenamefont {Beli{\'c}}, \citenamefont {Zhu},
  \citenamefont {Zhong}, \citenamefont {Zhang}, \citenamefont
  {Christodoulides},\ and\ \citenamefont {Xiao}}]{zhang2016pt}%
  \BibitemOpen
  \bibfield  {author} {\bibinfo {author} {\bibfnamefont {Y.}~\bibnamefont
  {Zhang}}, \bibinfo {author} {\bibfnamefont {H.}~\bibnamefont {Zhong}},
  \bibinfo {author} {\bibfnamefont {M.~R.}\ \bibnamefont {Beli{\'c}}}, \bibinfo
  {author} {\bibfnamefont {Y.}~\bibnamefont {Zhu}}, \bibinfo {author}
  {\bibfnamefont {W.}~\bibnamefont {Zhong}}, \bibinfo {author} {\bibfnamefont
  {Y.}~\bibnamefont {Zhang}}, \bibinfo {author} {\bibfnamefont {D.~N.}\
  \bibnamefont {Christodoulides}},\ and\ \bibinfo {author} {\bibfnamefont
  {M.}~\bibnamefont {Xiao}},\ }\href
  {https://doi.org/https://doi.org/10.1002/lpor.201600037} {\bibfield
  {journal} {\bibinfo  {journal} {Laser Photonics Rev.}\ }\textbf {\bibinfo
  {volume} {10}},\ \bibinfo {pages} {526} (\bibinfo {year} {2016})}\BibitemShut
  {NoStop}%
\bibitem [{\citenamefont {Heydari}\ and\ \citenamefont
  {Atangana}(2019)}]{heydari2019cardinal}%
  \BibitemOpen
  \bibfield  {author} {\bibinfo {author} {\bibfnamefont {M.}~\bibnamefont
  {Heydari}}\ and\ \bibinfo {author} {\bibfnamefont {A.}~\bibnamefont
  {Atangana}},\ }\href
  {https://doi.org/https://doi.org/10.1016/j.chaos.2019.08.009} {\bibfield
  {journal} {\bibinfo  {journal} {Chaos Solit. Fractals}\ }\textbf {\bibinfo
  {volume} {128}},\ \bibinfo {pages} {339} (\bibinfo {year}
  {2019})}\BibitemShut {NoStop}%
\bibitem [{\citenamefont {Stephanovich}\ \emph {et~al.}(2022)\citenamefont
  {Stephanovich}, \citenamefont {Kirichenko}, \citenamefont {Dugaev},
  \citenamefont {Sauco},\ and\ \citenamefont {Brito}}]{stephanovich2022}%
  \BibitemOpen
  \bibfield  {author} {\bibinfo {author} {\bibfnamefont {V.}~\bibnamefont
  {Stephanovich}}, \bibinfo {author} {\bibfnamefont {E.}~\bibnamefont
  {Kirichenko}}, \bibinfo {author} {\bibfnamefont {V.}~\bibnamefont {Dugaev}},
  \bibinfo {author} {\bibfnamefont {J.~H.}\ \bibnamefont {Sauco}},\ and\
  \bibinfo {author} {\bibfnamefont {B.~L.}\ \bibnamefont {Brito}},\ }\href
  {https://doi.org/doi.org/10.1038/s41598-022-16597-2} {\bibfield  {journal}
  {\bibinfo  {journal} {Scientific Reports}\ }\textbf {\bibinfo {volume}
  {12}},\ \bibinfo {pages} {12540} (\bibinfo {year} {2022})}\BibitemShut
  {NoStop}%
\bibitem [{\citenamefont {Gabrick}\ \emph {et~al.}(2023)\citenamefont
  {Gabrick}, \citenamefont {Sayari}, \citenamefont {{de Castro}}, \citenamefont
  {Trobia}, \citenamefont {Batista},\ and\ \citenamefont
  {Lenzi}}]{gabrick2023}%
  \BibitemOpen
  \bibfield  {author} {\bibinfo {author} {\bibfnamefont {E.}~\bibnamefont
  {Gabrick}}, \bibinfo {author} {\bibfnamefont {E.}~\bibnamefont {Sayari}},
  \bibinfo {author} {\bibfnamefont {A.}~\bibnamefont {{de Castro}}}, \bibinfo
  {author} {\bibfnamefont {J.}~\bibnamefont {Trobia}}, \bibinfo {author}
  {\bibfnamefont {A.}~\bibnamefont {Batista}},\ and\ \bibinfo {author}
  {\bibfnamefont {E.}~\bibnamefont {Lenzi}},\ }\href
  {https://doi.org/https://doi.org/10.1016/j.cnsns.2023.107275} {\bibfield
  {journal} {\bibinfo  {journal} {Commun. Nonlinear Sci. Numer. Simul.}\
  }\textbf {\bibinfo {volume} {123}},\ \bibinfo {pages} {107275} (\bibinfo
  {year} {2023})}\BibitemShut {NoStop}%
\bibitem [{\citenamefont {Lenzi}\ \emph {et~al.}(2023)\citenamefont {Lenzi},
  \citenamefont {Gabrick}, \citenamefont {Sayari}, \citenamefont {de~Castro},
  \citenamefont {Trobia},\ and\ \citenamefont {Batista}}]{lenzi2023}%
  \BibitemOpen
  \bibfield  {author} {\bibinfo {author} {\bibfnamefont {E.~K.}\ \bibnamefont
  {Lenzi}}, \bibinfo {author} {\bibfnamefont {E.~C.}\ \bibnamefont {Gabrick}},
  \bibinfo {author} {\bibfnamefont {E.}~\bibnamefont {Sayari}}, \bibinfo
  {author} {\bibfnamefont {A.~S.~M.}\ \bibnamefont {de~Castro}}, \bibinfo
  {author} {\bibfnamefont {J.}~\bibnamefont {Trobia}},\ and\ \bibinfo {author}
  {\bibfnamefont {A.~M.}\ \bibnamefont {Batista}},\ }\href
  {https://doi.org/10.3390/quantum5020029} {\bibfield  {journal} {\bibinfo
  {journal} {Quantum Reports}\ }\textbf {\bibinfo {volume} {5}},\ \bibinfo
  {pages} {442} (\bibinfo {year} {2023})}\BibitemShut {NoStop}%
\bibitem [{\citenamefont {Wu}\ \emph {et~al.}(2010)\citenamefont {Wu},
  \citenamefont {Huang}, \citenamefont {Cheng},\ and\ \citenamefont
  {Hsieh}}]{wu2010}%
  \BibitemOpen
  \bibfield  {author} {\bibinfo {author} {\bibfnamefont {J.-N.}\ \bibnamefont
  {Wu}}, \bibinfo {author} {\bibfnamefont {C.-H.}\ \bibnamefont {Huang}},
  \bibinfo {author} {\bibfnamefont {S.-C.}\ \bibnamefont {Cheng}},\ and\
  \bibinfo {author} {\bibfnamefont {W.-F.}\ \bibnamefont {Hsieh}},\ }\href
  {https://doi.org/10.1103/PhysRevA.81.023827} {\bibfield  {journal} {\bibinfo
  {journal} {Phys. Rev. A}\ }\textbf {\bibinfo {volume} {81}},\ \bibinfo
  {pages} {023827} (\bibinfo {year} {2010})}\BibitemShut {NoStop}%
\bibitem [{\citenamefont {Fujita}\ \emph {et~al.}(2005)\citenamefont {Fujita},
  \citenamefont {Takahashi}, \citenamefont {Tanaka}, \citenamefont {Asano},\
  and\ \citenamefont {Noda}}]{fujita2005}%
  \BibitemOpen
  \bibfield  {author} {\bibinfo {author} {\bibfnamefont {M.}~\bibnamefont
  {Fujita}}, \bibinfo {author} {\bibfnamefont {S.}~\bibnamefont {Takahashi}},
  \bibinfo {author} {\bibfnamefont {Y.}~\bibnamefont {Tanaka}}, \bibinfo
  {author} {\bibfnamefont {T.}~\bibnamefont {Asano}},\ and\ \bibinfo {author}
  {\bibfnamefont {S.}~\bibnamefont {Noda}},\ }\href
  {https://doi.org/10.1126/science.1110417} {\bibfield  {journal} {\bibinfo
  {journal} {Science}\ }\textbf {\bibinfo {volume} {308}},\ \bibinfo {pages}
  {1296} (\bibinfo {year} {2005})}\BibitemShut {NoStop}%
\bibitem [{\citenamefont {Laskin}(2000{\natexlab{a}})}]{laskin:00a}%
  \BibitemOpen
  \bibfield  {author} {\bibinfo {author} {\bibfnamefont {N.}~\bibnamefont
  {Laskin}},\ }\href
  {https://doi.org/https://doi.org/10.1016/S0375-9601(00)00201-2} {\bibfield
  {journal} {\bibinfo  {journal} {Phys. Lett. A}\ }\textbf {\bibinfo {volume}
  {268}},\ \bibinfo {pages} {298} (\bibinfo {year}
  {2000}{\natexlab{a}})}\BibitemShut {NoStop}%
\bibitem [{\citenamefont {Laskin}(2000{\natexlab{b}})}]{laskin:00b}%
  \BibitemOpen
  \bibfield  {author} {\bibinfo {author} {\bibfnamefont {N.}~\bibnamefont
  {Laskin}},\ }\href {https://doi.org/https://doi.org/10.1103/PhysRevE.62.3135}
  {\bibfield  {journal} {\bibinfo  {journal} {Phys. Rev. E}\ }\textbf {\bibinfo
  {volume} {62}},\ \bibinfo {pages} {3135} (\bibinfo {year}
  {2000}{\natexlab{b}})}\BibitemShut {NoStop}%
\bibitem [{\citenamefont {Wei}(2016)}]{wei:16}%
  \BibitemOpen
  \bibfield  {author} {\bibinfo {author} {\bibfnamefont {Y.}~\bibnamefont
  {Wei}},\ }\href {https://doi.org/https://doi.org/10.1103/PhysRevE.93.066103}
  {\bibfield  {journal} {\bibinfo  {journal} {Phys. Rev. E}\ }\textbf {\bibinfo
  {volume} {93}},\ \bibinfo {pages} {066103} (\bibinfo {year}
  {2016})}\BibitemShut {NoStop}%
\bibitem [{\citenamefont {Laskin}(2016)}]{laskin:16}%
  \BibitemOpen
  \bibfield  {author} {\bibinfo {author} {\bibfnamefont {N.}~\bibnamefont
  {Laskin}},\ }\href
  {https://doi.org/https://doi.org/10.1103/PhysRevE.93.066104} {\bibfield
  {journal} {\bibinfo  {journal} {Phys. Rev. E}\ }\textbf {\bibinfo {volume}
  {93}},\ \bibinfo {pages} {066104} (\bibinfo {year} {2016})}\BibitemShut
  {NoStop}%
\bibitem [{\citenamefont {Naber}(2004)}]{naber:04}%
  \BibitemOpen
  \bibfield  {author} {\bibinfo {author} {\bibfnamefont {M.}~\bibnamefont
  {Naber}},\ }\href {https://doi.org/https://doi.org/10.1063/1.1769611}
  {\bibfield  {journal} {\bibinfo  {journal} {J. Math. Phys.}\ }\textbf
  {\bibinfo {volume} {45}},\ \bibinfo {pages} {3339} (\bibinfo {year}
  {2004})}\BibitemShut {NoStop}%
\bibitem [{\citenamefont {Iomin}(2009)}]{iomin2009fractional}%
  \BibitemOpen
  \bibfield  {author} {\bibinfo {author} {\bibfnamefont {A.}~\bibnamefont
  {Iomin}},\ }\href
  {https://doi.org/https://doi.org/10.1103/PhysRevE.80.022103} {\bibfield
  {journal} {\bibinfo  {journal} {Phys. Rev. E}\ }\textbf {\bibinfo {volume}
  {80}},\ \bibinfo {pages} {022103} (\bibinfo {year} {2009})}\BibitemShut
  {NoStop}%
\bibitem [{\citenamefont {Lenzi}\ \emph {et~al.}(2013)\citenamefont {Lenzi},
  \citenamefont {Ribeiro}, \citenamefont {dos Santos}, \citenamefont
  {Rossato},\ and\ \citenamefont {Mendes}}]{lenzi2013time}%
  \BibitemOpen
  \bibfield  {author} {\bibinfo {author} {\bibfnamefont {E.~K.}\ \bibnamefont
  {Lenzi}}, \bibinfo {author} {\bibfnamefont {H.~V.}\ \bibnamefont {Ribeiro}},
  \bibinfo {author} {\bibfnamefont {M.~A.~F.}\ \bibnamefont {dos Santos}},
  \bibinfo {author} {\bibfnamefont {R.}~\bibnamefont {Rossato}},\ and\ \bibinfo
  {author} {\bibfnamefont {R.~S.}\ \bibnamefont {Mendes}},\ }\href
  {https://doi.org/https://doi.org/10.1063/1.4819253} {\bibfield  {journal}
  {\bibinfo  {journal} {J. Math. Phys.}\ }\textbf {\bibinfo {volume} {54}},\
  \bibinfo {pages} {082107} (\bibinfo {year} {2013})}\BibitemShut {NoStop}%
\bibitem [{\citenamefont {Iomin}(2020)}]{iomin2020}%
  \BibitemOpen
  \bibfield  {author} {\bibinfo {author} {\bibfnamefont {A.}~\bibnamefont
  {Iomin}},\ }\href
  {https://doi.org/https://doi.org/10.1016/j.chaos.2020.110305} {\bibfield
  {journal} {\bibinfo  {journal} {Chaos Solit. Fractals}\ }\textbf {\bibinfo
  {volume} {139}},\ \bibinfo {pages} {110305} (\bibinfo {year}
  {2020})}\BibitemShut {NoStop}%
\bibitem [{\citenamefont {Ekert}\ \emph {et~al.}(1992)\citenamefont {Ekert},
  \citenamefont {Rarity}, \citenamefont {Tapster},\ and\ \citenamefont
  {Massimo~Palma}}]{ekert1992}%
  \BibitemOpen
  \bibfield  {author} {\bibinfo {author} {\bibfnamefont {A.~K.}\ \bibnamefont
  {Ekert}}, \bibinfo {author} {\bibfnamefont {J.~G.}\ \bibnamefont {Rarity}},
  \bibinfo {author} {\bibfnamefont {P.~R.}\ \bibnamefont {Tapster}},\ and\
  \bibinfo {author} {\bibfnamefont {G.}~\bibnamefont {Massimo~Palma}},\ }\href
  {https://doi.org/10.1103/PhysRevLett.69.1293} {\bibfield  {journal} {\bibinfo
   {journal} {Phys. Rev. Lett.}\ }\textbf {\bibinfo {volume} {69}},\ \bibinfo
  {pages} {1293} (\bibinfo {year} {1992})}\BibitemShut {NoStop}%
\bibitem [{\citenamefont {Bennett}\ \emph {et~al.}(1993)\citenamefont
  {Bennett}, \citenamefont {Brassard}, \citenamefont {Cr\'epeau}, \citenamefont
  {Jozsa}, \citenamefont {Peres},\ and\ \citenamefont
  {Wootters}}]{bennett1993}%
  \BibitemOpen
  \bibfield  {author} {\bibinfo {author} {\bibfnamefont {C.~H.}\ \bibnamefont
  {Bennett}}, \bibinfo {author} {\bibfnamefont {G.}~\bibnamefont {Brassard}},
  \bibinfo {author} {\bibfnamefont {C.}~\bibnamefont {Cr\'epeau}}, \bibinfo
  {author} {\bibfnamefont {R.}~\bibnamefont {Jozsa}}, \bibinfo {author}
  {\bibfnamefont {A.}~\bibnamefont {Peres}},\ and\ \bibinfo {author}
  {\bibfnamefont {W.~K.}\ \bibnamefont {Wootters}},\ }\href
  {https://doi.org/10.1103/PhysRevLett.70.1895} {\bibfield  {journal} {\bibinfo
   {journal} {Phys. Rev. Lett.}\ }\textbf {\bibinfo {volume} {70}},\ \bibinfo
  {pages} {1895} (\bibinfo {year} {1993})}\BibitemShut {NoStop}%
\bibitem [{\citenamefont {Giovannetti}\ \emph {et~al.}(2011)\citenamefont
  {Giovannetti}, \citenamefont {Lloyd},\ and\ \citenamefont
  {Maccone}}]{giovannetti2011}%
  \BibitemOpen
  \bibfield  {author} {\bibinfo {author} {\bibfnamefont {V.}~\bibnamefont
  {Giovannetti}}, \bibinfo {author} {\bibfnamefont {S.}~\bibnamefont {Lloyd}},\
  and\ \bibinfo {author} {\bibfnamefont {L.}~\bibnamefont {Maccone}},\ }\href
  {https://doi.org/https://doi.org/10.1038/nphoton.2011.35} {\bibfield
  {journal} {\bibinfo  {journal} {Nat. Photonics}\ }\textbf {\bibinfo {volume}
  {5}},\ \bibinfo {pages} {222} (\bibinfo {year} {2011})}\BibitemShut {NoStop}%
\bibitem [{\citenamefont {Guha}\ \emph {et~al.}(2023)\citenamefont {Guha},
  \citenamefont {Roy},\ and\ \citenamefont {Chiribella}}]{guha2023}%
  \BibitemOpen
  \bibfield  {author} {\bibinfo {author} {\bibfnamefont {T.}~\bibnamefont
  {Guha}}, \bibinfo {author} {\bibfnamefont {S.}~\bibnamefont {Roy}},\ and\
  \bibinfo {author} {\bibfnamefont {G.}~\bibnamefont {Chiribella}},\ }\href
  {https://doi.org/10.1103/PhysRevResearch.5.033214} {\bibfield  {journal}
  {\bibinfo  {journal} {Phys. Rev. Res.}\ }\textbf {\bibinfo {volume} {5}},\
  \bibinfo {pages} {033214} (\bibinfo {year} {2023})}\BibitemShut {NoStop}%
\bibitem [{\citenamefont {Raussendorf}\ and\ \citenamefont
  {Briegel}(2001)}]{raussendorf2001}%
  \BibitemOpen
  \bibfield  {author} {\bibinfo {author} {\bibfnamefont {R.}~\bibnamefont
  {Raussendorf}}\ and\ \bibinfo {author} {\bibfnamefont {H.~J.}\ \bibnamefont
  {Briegel}},\ }\href {https://doi.org/10.1103/PhysRevLett.86.5188} {\bibfield
  {journal} {\bibinfo  {journal} {Phys. Rev. Lett.}\ }\textbf {\bibinfo
  {volume} {86}},\ \bibinfo {pages} {5188} (\bibinfo {year}
  {2001})}\BibitemShut {NoStop}%
\bibitem [{\citenamefont {Zu}\ \emph {et~al.}(2021)\citenamefont {Zu},
  \citenamefont {Gao},\ and\ \citenamefont {Yu}}]{zu2021}%
  \BibitemOpen
  \bibfield  {author} {\bibinfo {author} {\bibfnamefont {C.}~\bibnamefont
  {Zu}}, \bibinfo {author} {\bibfnamefont {Y.}~\bibnamefont {Gao}},\ and\
  \bibinfo {author} {\bibfnamefont {X.}~\bibnamefont {Yu}},\ }\href
  {https://doi.org/https://doi.org/10.1016/j.chaos.2021.110930} {\bibfield
  {journal} {\bibinfo  {journal} {Chaos, Solit. Fractals}\ }\textbf {\bibinfo
  {volume} {147}},\ \bibinfo {pages} {110930} (\bibinfo {year}
  {2021})}\BibitemShut {NoStop}%
\bibitem [{\citenamefont {Zu}\ and\ \citenamefont {Yu}(2022)}]{zu2022}%
  \BibitemOpen
  \bibfield  {author} {\bibinfo {author} {\bibfnamefont {C.}~\bibnamefont
  {Zu}}\ and\ \bibinfo {author} {\bibfnamefont {X.}~\bibnamefont {Yu}},\ }\href
  {https://doi.org/https://doi.org/10.1016/j.chaos.2022.111941} {\bibfield
  {journal} {\bibinfo  {journal} {Chaos, Solit. Fractals}\ }\textbf {\bibinfo
  {volume} {157}},\ \bibinfo {pages} {111941} (\bibinfo {year}
  {2022})}\BibitemShut {NoStop}%
\bibitem [{\citenamefont {Jaynes}\ and\ \citenamefont
  {Cummings}(1963)}]{jaynes1963}%
  \BibitemOpen
  \bibfield  {author} {\bibinfo {author} {\bibfnamefont {E.}~\bibnamefont
  {Jaynes}}\ and\ \bibinfo {author} {\bibfnamefont {F.}~\bibnamefont
  {Cummings}},\ }\href {https://doi.org/10.1109/PROC.1963.1664} {\bibfield
  {journal} {\bibinfo  {journal} {Proceedings of the IEEE}\ }\textbf {\bibinfo
  {volume} {51}},\ \bibinfo {pages} {89} (\bibinfo {year} {1963})}\BibitemShut
  {NoStop}%
\bibitem [{\citenamefont {Lu}\ and\ \citenamefont {Yu}(2017)}]{lu2017}%
  \BibitemOpen
  \bibfield  {author} {\bibinfo {author} {\bibfnamefont {L.}~\bibnamefont
  {Lu}}\ and\ \bibinfo {author} {\bibfnamefont {X.}~\bibnamefont {Yu}},\ }\href
  {https://doi.org/10.1088/1612-202X/aa8bc4} {\bibfield  {journal} {\bibinfo
  {journal} {Laser Phys. Lett.}\ }\textbf {\bibinfo {volume} {14}},\ \bibinfo
  {pages} {115202} (\bibinfo {year} {2017})}\BibitemShut {NoStop}%
\bibitem [{\citenamefont {Lu}\ and\ \citenamefont {Yu}(2018)}]{lu2018}%
  \BibitemOpen
  \bibfield  {author} {\bibinfo {author} {\bibfnamefont {L.}~\bibnamefont
  {Lu}}\ and\ \bibinfo {author} {\bibfnamefont {X.}~\bibnamefont {Yu}},\ }\href
  {https://doi.org/https://doi.org/10.1016/j.aop.2018.03.017} {\bibfield
  {journal} {\bibinfo  {journal} {Ann. Phys.}\ }\textbf {\bibinfo {volume}
  {392}},\ \bibinfo {pages} {260} (\bibinfo {year} {2018})}\BibitemShut
  {NoStop}%
\bibitem [{\citenamefont {Cius}\ \emph
  {et~al.}(2022{\natexlab{a}})\citenamefont {Cius}, \citenamefont {Menon},
  \citenamefont {dos Santos}, \citenamefont {de~Castro},\ and\ \citenamefont
  {Andrade}}]{cius22frac}%
  \BibitemOpen
  \bibfield  {author} {\bibinfo {author} {\bibfnamefont {D.}~\bibnamefont
  {Cius}}, \bibinfo {author} {\bibfnamefont {L.}~\bibnamefont {Menon}},
  \bibinfo {author} {\bibfnamefont {M.~A.~F.}\ \bibnamefont {dos Santos}},
  \bibinfo {author} {\bibfnamefont {A.~S.~M.}\ \bibnamefont {de~Castro}},\ and\
  \bibinfo {author} {\bibfnamefont {F.~M.}\ \bibnamefont {Andrade}},\ }\href
  {https://doi.org/10.1103/PhysRevE.106.054126} {\bibfield  {journal} {\bibinfo
   {journal} {Phys. Rev. E}\ }\textbf {\bibinfo {volume} {106}},\ \bibinfo
  {pages} {054126} (\bibinfo {year} {2022}{\natexlab{a}})}\BibitemShut
  {NoStop}%
\bibitem [{\citenamefont {Fring}\ and\ \citenamefont
  {Moussa}(2016{\natexlab{a}})}]{fring:16a}%
  \BibitemOpen
  \bibfield  {author} {\bibinfo {author} {\bibfnamefont {A.}~\bibnamefont
  {Fring}}\ and\ \bibinfo {author} {\bibfnamefont {M.~H.~Y.}\ \bibnamefont
  {Moussa}},\ }\href
  {https://doi.org/https://doi.org/10.1103/PhysRevA.93.042114} {\bibfield
  {journal} {\bibinfo  {journal} {Phys. Rev. A}\ }\textbf {\bibinfo {volume}
  {93}},\ \bibinfo {pages} {042114} (\bibinfo {year}
  {2016}{\natexlab{a}})}\BibitemShut {NoStop}%
\bibitem [{\citenamefont {Fring}\ and\ \citenamefont {Frith}(2017)}]{fring:17}%
  \BibitemOpen
  \bibfield  {author} {\bibinfo {author} {\bibfnamefont {A.}~\bibnamefont
  {Fring}}\ and\ \bibinfo {author} {\bibfnamefont {T.}~\bibnamefont {Frith}},\
  }\href {https://doi.org/https://doi.org/10.1016/j.physleta.2017.05.041}
  {\bibfield  {journal} {\bibinfo  {journal} {Phys. Lett. A}\ }\textbf
  {\bibinfo {volume} {381}},\ \bibinfo {pages} {2318} (\bibinfo {year}
  {2017})}\BibitemShut {NoStop}%
\bibitem [{\citenamefont {Laskin}(2017)}]{laskin2017time}%
  \BibitemOpen
  \bibfield  {author} {\bibinfo {author} {\bibfnamefont {N.}~\bibnamefont
  {Laskin}},\ }\href
  {https://doi.org/https://doi.org/10.1016/j.chaos.2017.04.010} {\bibfield
  {journal} {\bibinfo  {journal} {Chaos Solit. Fractals}\ }\textbf {\bibinfo
  {volume} {102}},\ \bibinfo {pages} {16} (\bibinfo {year} {2017})}\BibitemShut
  {NoStop}%
\bibitem [{\citenamefont {Iomin}(2019{\natexlab{a}})}]{iomin2019app}%
  \BibitemOpen
  \bibfield  {author} {\bibinfo {author} {\bibfnamefont {A.}~\bibnamefont
  {Iomin}},\ }\bibinfo {title} {Fractional time quantum mechanics},\ in\ \href
  {https://doi.org/doi:10.1515/9783110571721-013} {\emph {\bibinfo {booktitle}
  {Volume 5 Applications in Physics, Part B}}},\ \bibinfo {editor} {edited by\
  \bibinfo {editor} {\bibfnamefont {V.~E.}\ \bibnamefont {Tarasov}}}\ (\bibinfo
   {publisher} {De Gruyter},\ \bibinfo {address} {Berlin, Boston},\ \bibinfo
  {year} {2019})\ pp.\ \bibinfo {pages} {299--316}\BibitemShut {NoStop}%
\bibitem [{\citenamefont {Zhang}\ \emph {et~al.}(2021)\citenamefont {Zhang},
  \citenamefont {Yang}, \citenamefont {Wei},\ and\ \citenamefont
  {Luo}}]{zhang2021quantization}%
  \BibitemOpen
  \bibfield  {author} {\bibinfo {author} {\bibfnamefont {X.}~\bibnamefont
  {Zhang}}, \bibinfo {author} {\bibfnamefont {B.}~\bibnamefont {Yang}},
  \bibinfo {author} {\bibfnamefont {C.}~\bibnamefont {Wei}},\ and\ \bibinfo
  {author} {\bibfnamefont {M.}~\bibnamefont {Luo}},\ }\href
  {https://doi.org/https://doi.org/10.1016/j.cnsns.2020.105531} {\bibfield
  {journal} {\bibinfo  {journal} {Commun. Nonlinear Sci. Numer. Simul.}\
  }\textbf {\bibinfo {volume} {93}},\ \bibinfo {pages} {105531} (\bibinfo
  {year} {2021})}\BibitemShut {NoStop}%
\bibitem [{\citenamefont {Iomin}(2019{\natexlab{b}})}]{iomin2019fractional}%
  \BibitemOpen
  \bibfield  {author} {\bibinfo {author} {\bibfnamefont {A.}~\bibnamefont
  {Iomin}},\ }\href
  {https://doi.org/https://doi.org/10.1016/j.csfx.2018.100001} {\bibfield
  {journal} {\bibinfo  {journal} {Chaos Solit. Fractals: X}\ }\textbf {\bibinfo
  {volume} {1}},\ \bibinfo {pages} {100001} (\bibinfo {year}
  {2019}{\natexlab{b}})}\BibitemShut {NoStop}%
\bibitem [{\citenamefont {Luiz}\ \emph {et~al.}(2020)\citenamefont {Luiz},
  \citenamefont {de~Ponte},\ and\ \citenamefont {Moussa}}]{luiz:20}%
  \BibitemOpen
  \bibfield  {author} {\bibinfo {author} {\bibfnamefont {F.~S.}\ \bibnamefont
  {Luiz}}, \bibinfo {author} {\bibfnamefont {M.~A.}\ \bibnamefont {de~Ponte}},\
  and\ \bibinfo {author} {\bibfnamefont {M.~H.~Y.}\ \bibnamefont {Moussa}},\
  }\href {https://doi.org/https://doi.org/10.1088/1402-4896/ab80e5} {\bibfield
  {journal} {\bibinfo  {journal} {Phys. Scr.}\ }\textbf {\bibinfo {volume}
  {95}},\ \bibinfo {pages} {065211} (\bibinfo {year} {2020})}\BibitemShut
  {NoStop}%
\bibitem [{\citenamefont {Fring}\ and\ \citenamefont
  {Moussa}(2016{\natexlab{b}})}]{fring16b}%
  \BibitemOpen
  \bibfield  {author} {\bibinfo {author} {\bibfnamefont {A.}~\bibnamefont
  {Fring}}\ and\ \bibinfo {author} {\bibfnamefont {M.~H.~Y.}\ \bibnamefont
  {Moussa}},\ }\href {https://doi.org/10.1103/PhysRevA.94.042128} {\bibfield
  {journal} {\bibinfo  {journal} {Phys. Rev. A}\ }\textbf {\bibinfo {volume}
  {94}},\ \bibinfo {pages} {042128} (\bibinfo {year}
  {2016}{\natexlab{b}})}\BibitemShut {NoStop}%
\bibitem [{\citenamefont {Fring}\ and\ \citenamefont {Frith}(2019)}]{fring:19}%
  \BibitemOpen
  \bibfield  {author} {\bibinfo {author} {\bibfnamefont {A.}~\bibnamefont
  {Fring}}\ and\ \bibinfo {author} {\bibfnamefont {T.}~\bibnamefont {Frith}},\
  }\href {https://doi.org/https://doi.org/10.1103/PhysRevA.100.010102}
  {\bibfield  {journal} {\bibinfo  {journal} {Phys. Rev. A}\ }\textbf {\bibinfo
  {volume} {100}},\ \bibinfo {pages} {010102} (\bibinfo {year}
  {2019})}\BibitemShut {NoStop}%
\bibitem [{\citenamefont {Mana}\ \emph {et~al.}(2020)\citenamefont {Mana},
  \citenamefont {Zaidi},\ and\ \citenamefont {Maamache}}]{mana:20}%
  \BibitemOpen
  \bibfield  {author} {\bibinfo {author} {\bibfnamefont {N.}~\bibnamefont
  {Mana}}, \bibinfo {author} {\bibfnamefont {O.}~\bibnamefont {Zaidi}},\ and\
  \bibinfo {author} {\bibfnamefont {M.}~\bibnamefont {Maamache}},\ }\href
  {https://doi.org/https://doi.org/10.1063/5.0013723} {\bibfield  {journal}
  {\bibinfo  {journal} {J. Math. Phys.}\ }\textbf {\bibinfo {volume} {61}},\
  \bibinfo {pages} {102103} (\bibinfo {year} {2020})}\BibitemShut {NoStop}%
\bibitem [{\citenamefont {Koussa}\ \emph {et~al.}(2020)\citenamefont {Koussa},
  \citenamefont {Attia},\ and\ \citenamefont {Maamache}}]{koussa:20}%
  \BibitemOpen
  \bibfield  {author} {\bibinfo {author} {\bibfnamefont {W.}~\bibnamefont
  {Koussa}}, \bibinfo {author} {\bibfnamefont {M.}~\bibnamefont {Attia}},\ and\
  \bibinfo {author} {\bibfnamefont {M.}~\bibnamefont {Maamache}},\ }\href
  {https://doi.org/https://doi.org/10.1063/1.5145269} {\bibfield  {journal}
  {\bibinfo  {journal} {J. Math. Phys.}\ }\textbf {\bibinfo {volume} {61}},\
  \bibinfo {pages} {042101} (\bibinfo {year} {2020})}\BibitemShut {NoStop}%
\bibitem [{\citenamefont {Cius}\ \emph
  {et~al.}(2022{\natexlab{b}})\citenamefont {Cius}, \citenamefont {Andrade},
  \citenamefont {{de Castro}},\ and\ \citenamefont {Moussa}}]{cius:22}%
  \BibitemOpen
  \bibfield  {author} {\bibinfo {author} {\bibfnamefont {D.}~\bibnamefont
  {Cius}}, \bibinfo {author} {\bibfnamefont {F.}~\bibnamefont {Andrade}},
  \bibinfo {author} {\bibfnamefont {A.}~\bibnamefont {{de Castro}}},\ and\
  \bibinfo {author} {\bibfnamefont {M.}~\bibnamefont {Moussa}},\ }\href
  {https://doi.org/https://doi.org/10.1016/j.physa.2022.126945} {\bibfield
  {journal} {\bibinfo  {journal} {Phys. A: Stat. Mech. Appl.}\ }\textbf
  {\bibinfo {volume} {593}},\ \bibinfo {pages} {126945} (\bibinfo {year}
  {2022}{\natexlab{b}})}\BibitemShut {NoStop}%
\bibitem [{\citenamefont {Cius}\ \emph {et~al.}(2023)\citenamefont {Cius},
  \citenamefont {Uhdre}, \citenamefont {de~Castro},\ and\ \citenamefont
  {Andrade}}]{cius2023}%
  \BibitemOpen
  \bibfield  {author} {\bibinfo {author} {\bibfnamefont {D.}~\bibnamefont
  {Cius}}, \bibinfo {author} {\bibfnamefont {G.~M.}\ \bibnamefont {Uhdre}},
  \bibinfo {author} {\bibfnamefont {A.~S.~M.}\ \bibnamefont {de~Castro}},\ and\
  \bibinfo {author} {\bibfnamefont {F.~M.}\ \bibnamefont {Andrade}},\ }\href
  {https://doi.org/10.1103/PhysRevA.107.022403} {\bibfield  {journal} {\bibinfo
   {journal} {Phys. Rev. A}\ }\textbf {\bibinfo {volume} {107}},\ \bibinfo
  {pages} {022403} (\bibinfo {year} {2023})}\BibitemShut {NoStop}%
\bibitem [{\citenamefont {Larson}\ and\ \citenamefont
  {Mavrogordatos}(2021)}]{larson2021}%
  \BibitemOpen
  \bibfield  {author} {\bibinfo {author} {\bibfnamefont {J.}~\bibnamefont
  {Larson}}\ and\ \bibinfo {author} {\bibfnamefont {T.}~\bibnamefont
  {Mavrogordatos}},\ }\href {https://doi.org/10.1088/978-0-7503-3447-1} {\emph
  {\bibinfo {title} {The Jaynes–Cummings Model and Its Descendants}}},\
  2053-2563\ (\bibinfo  {publisher} {IOP Publishing},\ \bibinfo {year}
  {2021})\BibitemShut {NoStop}%
\bibitem [{\citenamefont {Iomin}(2024)}]{iomin24}%
  \BibitemOpen
  \bibfield  {author} {\bibinfo {author} {\bibfnamefont {A.}~\bibnamefont
  {Iomin}},\ }\href {https://doi.org/10.1063/5.0226335} {\bibfield  {journal}
  {\bibinfo  {journal} {Chaos: An Interdisciplinary Journal of Nonlinear
  Science}\ }\textbf {\bibinfo {volume} {34}},\ \bibinfo {pages} {093135}
  (\bibinfo {year} {2024})}\BibitemShut {NoStop}%
\bibitem [{\citenamefont {Dias}\ and\ \citenamefont {Parisio}(2017)}]{dias17}%
  \BibitemOpen
  \bibfield  {author} {\bibinfo {author} {\bibfnamefont {E.~O.}\ \bibnamefont
  {Dias}}\ and\ \bibinfo {author} {\bibfnamefont {F.}~\bibnamefont {Parisio}},\
  }\href {https://doi.org/10.1103/PhysRevA.95.032133} {\bibfield  {journal}
  {\bibinfo  {journal} {Phys. Rev. A}\ }\textbf {\bibinfo {volume} {95}},\
  \bibinfo {pages} {032133} (\bibinfo {year} {2017})}\BibitemShut {NoStop}%
\bibitem [{\citenamefont {Beims}\ and\ \citenamefont
  {de~Lara}(2024)}]{beims24}%
  \BibitemOpen
  \bibfield  {author} {\bibinfo {author} {\bibfnamefont {M.~W.}\ \bibnamefont
  {Beims}}\ and\ \bibinfo {author} {\bibfnamefont {A.~J.}\ \bibnamefont
  {de~Lara}},\ }\href
  {https://doi.org/https://doi.org/10.1007/s10701-024-00787-1} {\bibfield
  {journal} {\bibinfo  {journal} {Foundations of Physics}\ }\textbf {\bibinfo
  {volume} {54}},\ \bibinfo {pages} {55} (\bibinfo {year} {2024})}\BibitemShut
  {NoStop}%
\bibitem [{\citenamefont {Tarasov}(2010)}]{tarasov2010}%
  \BibitemOpen
  \bibfield  {author} {\bibinfo {author} {\bibfnamefont {V.~E.}\ \bibnamefont
  {Tarasov}},\ }\bibinfo {title} {Fractional {D}ynamics of {O}pen {Q}uantum
  {S}ystems},\ in\ \href {https://doi.org/10.1007/978-3-642-14003-7_20} {\emph
  {\bibinfo {booktitle} {Fractional Dynamics}}}\ (\bibinfo  {publisher}
  {Springer Berlin Heidelberg},\ \bibinfo {address} {Berlin, Heidelberg},\
  \bibinfo {year} {2010})\ pp.\ \bibinfo {pages} {467--490}\BibitemShut
  {NoStop}%
\bibitem [{\citenamefont {Dutra}\ \emph {et~al.}(1994)\citenamefont {Dutra},
  \citenamefont {Knight},\ and\ \citenamefont {Moya-Cessa}}]{dutra1993}%
  \BibitemOpen
  \bibfield  {author} {\bibinfo {author} {\bibfnamefont {S.~M.}\ \bibnamefont
  {Dutra}}, \bibinfo {author} {\bibfnamefont {P.~L.}\ \bibnamefont {Knight}},\
  and\ \bibinfo {author} {\bibfnamefont {H.}~\bibnamefont {Moya-Cessa}},\
  }\href {https://doi.org/10.1103/PhysRevA.49.1993} {\bibfield  {journal}
  {\bibinfo  {journal} {Phys. Rev. A}\ }\textbf {\bibinfo {volume} {49}},\
  \bibinfo {pages} {1993} (\bibinfo {year} {1994})}\BibitemShut {NoStop}%
\bibitem [{\citenamefont {Schr{\"o}dinger}(1935)}]{schrodinger1935}%
  \BibitemOpen
  \bibfield  {author} {\bibinfo {author} {\bibfnamefont {E.}~\bibnamefont
  {Schr{\"o}dinger}},\ }\href
  {https://doi.org/https://doi.org/10.1007/BF01491891} {\bibfield  {journal}
  {\bibinfo  {journal} {Naturwissenschaften}\ }\textbf {\bibinfo {volume}
  {23}},\ \bibinfo {pages} {844} (\bibinfo {year} {1935})}\BibitemShut
  {NoStop}%
\bibitem [{\citenamefont {Horodecki}\ \emph {et~al.}(2009)\citenamefont
  {Horodecki}, \citenamefont {Horodecki}, \citenamefont {Horodecki},\ and\
  \citenamefont {Horodecki}}]{horodecki2009}%
  \BibitemOpen
  \bibfield  {author} {\bibinfo {author} {\bibfnamefont {R.}~\bibnamefont
  {Horodecki}}, \bibinfo {author} {\bibfnamefont {P.}~\bibnamefont
  {Horodecki}}, \bibinfo {author} {\bibfnamefont {M.}~\bibnamefont
  {Horodecki}},\ and\ \bibinfo {author} {\bibfnamefont {K.}~\bibnamefont
  {Horodecki}},\ }\href {https://doi.org/10.1103/RevModPhys.81.865} {\bibfield
  {journal} {\bibinfo  {journal} {Rev. Mod. Phys.}\ }\textbf {\bibinfo {volume}
  {81}},\ \bibinfo {pages} {865} (\bibinfo {year} {2009})}\BibitemShut
  {NoStop}%
\bibitem [{\citenamefont {Werner}(1989)}]{werner1989}%
  \BibitemOpen
  \bibfield  {author} {\bibinfo {author} {\bibfnamefont {R.~F.}\ \bibnamefont
  {Werner}},\ }\href {https://doi.org/10.1103/PhysRevA.40.4277} {\bibfield
  {journal} {\bibinfo  {journal} {Phys. Rev. A}\ }\textbf {\bibinfo {volume}
  {40}},\ \bibinfo {pages} {4277} (\bibinfo {year} {1989})}\BibitemShut
  {NoStop}%
\bibitem [{\citenamefont {Einstein}\ \emph {et~al.}(1935)\citenamefont
  {Einstein}, \citenamefont {Podolsky},\ and\ \citenamefont
  {Rosen}}]{einstein1935}%
  \BibitemOpen
  \bibfield  {author} {\bibinfo {author} {\bibfnamefont {A.}~\bibnamefont
  {Einstein}}, \bibinfo {author} {\bibfnamefont {B.}~\bibnamefont {Podolsky}},\
  and\ \bibinfo {author} {\bibfnamefont {N.}~\bibnamefont {Rosen}},\ }\href
  {https://doi.org/10.1103/PhysRev.47.777} {\bibfield  {journal} {\bibinfo
  {journal} {Phys. Rev.}\ }\textbf {\bibinfo {volume} {47}},\ \bibinfo {pages}
  {777} (\bibinfo {year} {1935})}\BibitemShut {NoStop}%
\bibitem [{\citenamefont {Bell}(1964)}]{bell1964}%
  \BibitemOpen
  \bibfield  {author} {\bibinfo {author} {\bibfnamefont {J.~S.}\ \bibnamefont
  {Bell}},\ }\href {https://doi.org/10.1103/PhysicsPhysiqueFizika.1.195}
  {\bibfield  {journal} {\bibinfo  {journal} {Phys. Phys. Fiz.}\ }\textbf
  {\bibinfo {volume} {1}},\ \bibinfo {pages} {195} (\bibinfo {year}
  {1964})}\BibitemShut {NoStop}%
\bibitem [{\citenamefont {Petz}(2001)}]{petz2001}%
  \BibitemOpen
  \bibfield  {author} {\bibinfo {author} {\bibfnamefont {D.}~\bibnamefont
  {Petz}},\ }\bibinfo {title} {Entropy, von {N}eumann and the von {N}eumann
  {E}ntropy},\ in\ \href {https://doi.org/10.1007/978-94-017-2012-0_7} {\emph
  {\bibinfo {booktitle} {John von {N}eumann and the Foundations of Quantum
  Physics}}},\ \bibinfo {editor} {edited by\ \bibinfo {editor} {\bibfnamefont
  {M.}~\bibnamefont {R{\'e}dei}}\ and\ \bibinfo {editor} {\bibfnamefont
  {M.}~\bibnamefont {St{\"o}ltzner}}}\ (\bibinfo  {publisher} {Springer
  Netherlands},\ \bibinfo {address} {Dordrecht},\ \bibinfo {year} {2001})\ pp.\
  \bibinfo {pages} {83--96}\BibitemShut {NoStop}%
\bibitem [{\citenamefont {Phoenix}\ and\ \citenamefont
  {Knight}(1988)}]{phoenix1988}%
  \BibitemOpen
  \bibfield  {author} {\bibinfo {author} {\bibfnamefont {S.}~\bibnamefont
  {Phoenix}}\ and\ \bibinfo {author} {\bibfnamefont {P.}~\bibnamefont
  {Knight}},\ }\href
  {https://doi.org/https://doi.org/10.1016/0003-4916(88)90006-1} {\bibfield
  {journal} {\bibinfo  {journal} {Annals of Physics}\ }\textbf {\bibinfo
  {volume} {186}},\ \bibinfo {pages} {381} (\bibinfo {year}
  {1988})}\BibitemShut {NoStop}%
\bibitem [{\citenamefont {Phoenix}\ and\ \citenamefont
  {Knight}(1991)}]{phoenix1991s}%
  \BibitemOpen
  \bibfield  {author} {\bibinfo {author} {\bibfnamefont {S.~J.~D.}\
  \bibnamefont {Phoenix}}\ and\ \bibinfo {author} {\bibfnamefont {P.~L.}\
  \bibnamefont {Knight}},\ }\href {https://doi.org/10.1103/PhysRevA.44.6023}
  {\bibfield  {journal} {\bibinfo  {journal} {Phys. Rev. A}\ }\textbf {\bibinfo
  {volume} {44}},\ \bibinfo {pages} {6023} (\bibinfo {year}
  {1991})}\BibitemShut {NoStop}%
\bibitem [{\citenamefont {Yang}\ \emph {et~al.}(2006)\citenamefont {Yang},
  \citenamefont {Ya-Ping},\ and\ \citenamefont {Hong}}]{sun2006}%
  \BibitemOpen
  \bibfield  {author} {\bibinfo {author} {\bibfnamefont {S.}~\bibnamefont
  {Yang}}, \bibinfo {author} {\bibfnamefont {Y.}~\bibnamefont {Ya-Ping}},\ and\
  \bibinfo {author} {\bibfnamefont {C.}~\bibnamefont {Hong}},\ }\href
  {https://doi.org/10.1088/0256-307X/23/5/020} {\bibfield  {journal} {\bibinfo
  {journal} {Chin. Phys. Lett.}\ }\textbf {\bibinfo {volume} {23}},\ \bibinfo
  {pages} {1136} (\bibinfo {year} {2006})}\BibitemShut {NoStop}%
\bibitem [{\citenamefont {Uhdre}\ \emph {et~al.}(2022)\citenamefont {Uhdre},
  \citenamefont {Cius},\ and\ \citenamefont {Andrade}}]{uhdre2022}%
  \BibitemOpen
  \bibfield  {author} {\bibinfo {author} {\bibfnamefont {G.~M.}\ \bibnamefont
  {Uhdre}}, \bibinfo {author} {\bibfnamefont {D.}~\bibnamefont {Cius}},\ and\
  \bibinfo {author} {\bibfnamefont {F.~M.}\ \bibnamefont {Andrade}},\ }\href
  {https://doi.org/10.1103/PhysRevA.105.013703} {\bibfield  {journal} {\bibinfo
   {journal} {Phys. Rev. A}\ }\textbf {\bibinfo {volume} {105}},\ \bibinfo
  {pages} {013703} (\bibinfo {year} {2022})}\BibitemShut {NoStop}%
\end{thebibliography}%

\end{document}